\documentclass[
reprint,
superscriptaddress,
amsmath,amssymb,
aps,
prb,
twocolumn
]{revtex4-2}

\usepackage{graphicx}
\usepackage{dcolumn}
\usepackage{bm}
\usepackage[breaklinks]{hyperref}
\hypersetup{colorlinks=true, linkcolor=blue, citecolor=blue, filecolor=blue, urlcolor=blue}
\graphicspath{{figures}}
\usepackage{bm}

\usepackage{graphicx,subfigure}
\usepackage{epsfig}
\usepackage{bm}
\usepackage{dcolumn}
\usepackage{color}
\usepackage{physics}
\usepackage{float}
\makeatletter
\newcommand*{\rom}[1]{\expandafter\@slowromancap\romannumeral #1@}
\makeatother
\usepackage{lipsum}
\usepackage{subfigure}
\usepackage{dsfont}
\usepackage{enumitem}
\usepackage{tikz}
\usepackage[normalem]{ulem}
 
\begin{document}
\title{Resolving topological order crossovers on NISQ hardware}

\author{Ruizhe Shen}
\email{e0554228@u.nus.edu}
\affiliation{Department of Physics, National University of Singapore, Singapore 117551}

\author{Yin Zhong}
\email{zhongy@lzu.edu.cn}
\affiliation{School of Physical Science and Technology $\&$ Key Laboratory for Magnetism and Magnetic Materials of the MoE, Lanzhou University, Lanzhou 730000, China} %
\affiliation{Lanzhou Center for Theoretical Physics, Key Laboratory of Theoretical Physics of Gansu Province, Lanzhou 730000, China}
\author{Ching Hua Lee}
\email{phylch@nus.edu.sg}
\affiliation{Department of Physics, National University of Singapore, Singapore 117551}
\date{\today}

\begin{abstract}
Topological phases of matter provide a promising route toward robust quantum information processing, but on present-day noisy intermediate-scale quantum devices the experimentally relevant question is whether signatures of topological crossovers remain resolvable under realistic imperfections. Here, we address this question in the Wen--plaquette model through a two-stage strategy on the IBM Quantum hardware. We first use a tractable system to systematically characterize crossover signatures. Using variationally compiled equilibrium and quench-generated states, we resolve crossovers between stabilizer-dominated and trivial or disorder-dominated regimes through local plaquette stabilizers and a Wilson loop, and quantify their robustness against static disorder, deliberately amplified circuit noise, and effective non-Hermitian fields. The quench dynamics further reveal that plaquette-sector signatures remain substantially more stable deep in the strong-stabilizer regime than near the finite-size crossover. Building on the properties established in the small system, we extend the implementation to a physical two-dimensional IBM processor using a layered representative-qubit construction. The resulting lattice-averaged plaquette response exhibits only weak degradation under intentionally amplified local coherent perturbations. Together, these results connect controlled finite-size characterization with a scalable hardware implementation, providing a practical route for preparing and probing topological signatures on near-term quantum processors.
\end{abstract}

\pacs{}  
 \maketitle
\section{Introduction}\label{sec0}

Understanding quantum phases that cannot be characterized solely by local order parameters and spontaneous symmetry breaking remains a central challenge in condensed-matter physics \cite{girvin2019modern,marder2010condensed,beekman2019introduction,wen1990topological,chen2010local}. This broader perspective was strongly stimulated by landmark discoveries in strongly correlated quantum matter, most notably the fractional quantum Hall effect (FQHE) \cite{tsui1982two,laughlin1983anomalous,stormer1999fractional,jain1990theory,moore1991nonabelions,murthy2003hamiltonian} and high-temperature superconductivity in the cuprates \cite{bednorz1986possible,anderson1987resonating,anderson2004physics,taillefer2010scattering}. These developments exposed the limitations of descriptions based only on weakly interacting quasiparticles and conventional local order, and motivated the search for more general organizing principles for quantum matter. In particular, the FQHE established topological order as a fundamentally new paradigm characterized by long-range entanglement, topology-dependent ground-state degeneracy, fractionalized anyonic excitations, and nonlocal observables rather than a conventional symmetry-breaking order parameter \cite{wen1990topological,kitaev2003fault,levin2006detecting,kitaev2006topological,chen2010local,jiang2012identifying,nayak2008non,johnson2022classification}. Within the quantum Hall setting, model wave functions and their parent Hamiltonians further revealed how clustering, pseudopotential, and geometric principles encode strongly correlated topological states \cite{bernevig2008generalized,estienne2010clustering,lee2014lattice,lee2015geometric}. Beyond their fundamental significance, topologically ordered phases provide a natural foundation for robust quantum-information processing because quantum information may be stored nonlocally and manipulated through the fusion and braiding of anyonic excitations \cite{kitaev2003fault,dennis2002topological,nayak2008non,raussendorf2007topological,bombin2023logical}. Rapid advances in programmable quantum platforms have now enabled the preparation, detection, and manipulation of increasingly complex topological states, including toric-code and spin-liquid states, measurement- and feedforward-assisted preparation, and Abelian and non-Abelian anyonic processes on superconducting, trapped-ion, and neutral-atom processors \cite{satzinger2021realizing,semeghini2021probing,chen2024direct,google2023non,tantivasadakarn2023hierarchy,iqbal2024measurement,iqbal2024non,xu2024non,shen2025robust,ng2026digital}. More recent studies have extended these capabilities to qutrit topological codes, non-equilibrium topological order, digital simulations of the Kitaev honeycomb model, and the preparation of a fermionic Laughlin state \cite{iqbal2025qutrit,will2025probing,evered2025probing,shen2026laughlin}.

\begin{figure*}
	\centering
	\includegraphics[width=0.8\linewidth]{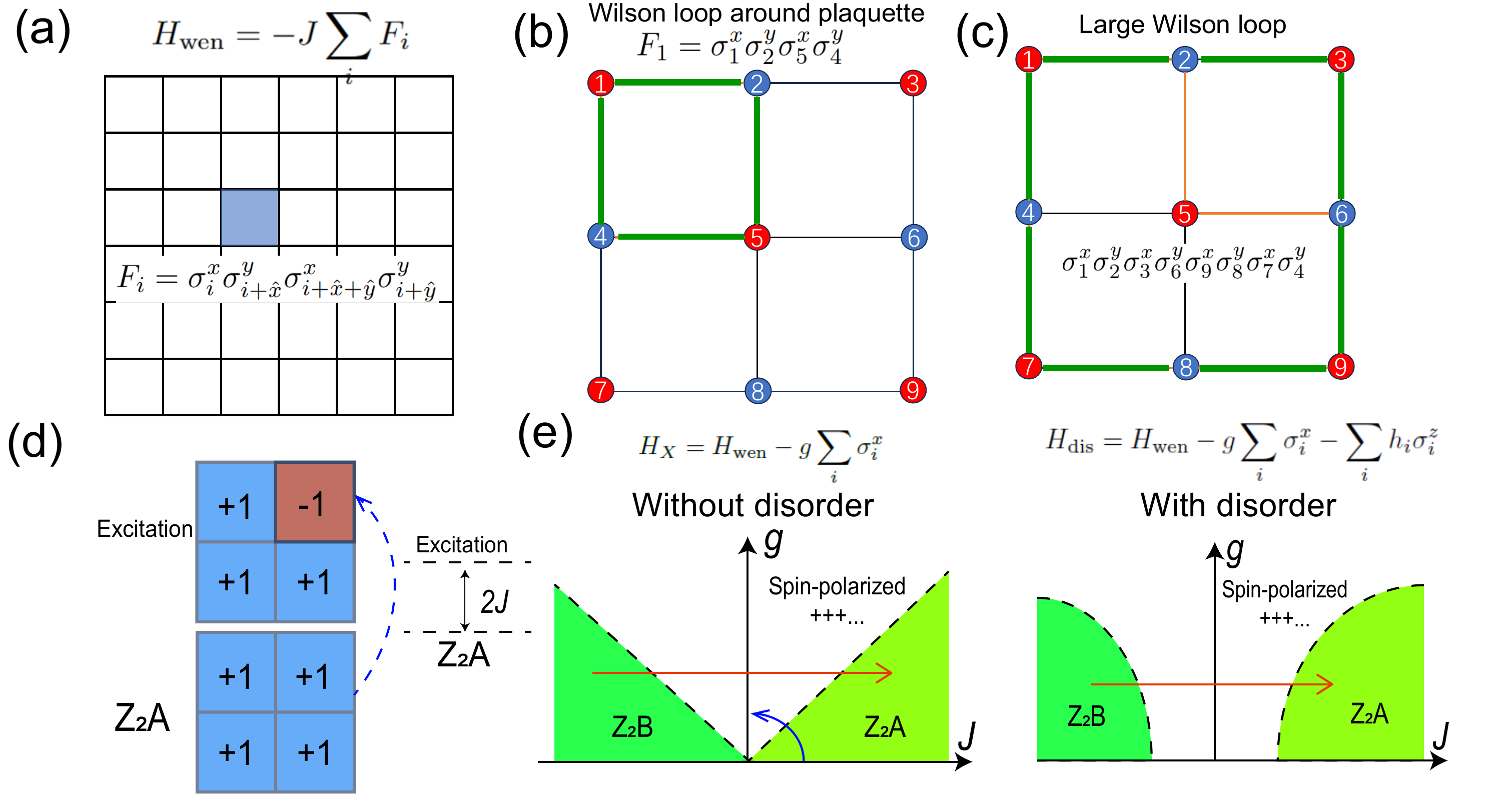}
	\caption{\textbf{Illustration of the Wen--plaquette model and its finite-size diagnostic structure.} \textbf{(a)} Schematic of the Wen--plaquette model [Eq.~\ref{wen}] on a square lattice. \textbf{(b,c)} Schematic definitions of the elementary plaquette operator [$F_1$, Eq.~\eqref{f}] and the extended Wilson-loop operator [Eq.~\eqref{w}], respectively. The latter corresponds to the nonlocal boundary string $W$. \textbf{(d)} Local excitation of the Wen--plaquette Hamiltonian: flipping a single plaquette eigenvalue from $+1$ to $-1$ creates a localized defect separated from the reference stabilizer sector by an energy cost $2J$. \textbf{(e)} Schematic finite-size response maps of the Wen--plaquette model in the clean case (left, Eq.~\ref{H}) and in the presence of disorder (right, Eq.~\ref{dis}). In both panels, the red arrows indicate the trajectory used to probe the crossover between the two stabilizer-dominated regimes, distinguished by the sign of the plaquette expectation value. In the clean system, the blue arrow indicates the trajectory used to probe the crossover from a regime with a strong Wilson-loop response to a transverse-field-dominated trivial regime.}
	\label{fig:figure1}
\end{figure*}

Topological order defines a class of quantum phases characterized by long-range entanglement and fundamentally distinct from conventional symmetry-breaking phases \cite{wen1990topological,levin2006detecting,kitaev2006topological,chen2010local,jiang2012identifying,johnson2022classification}. A central hallmark is the presence of ground-state degeneracy determined by global topological properties and stable against local perturbations. This nonlocal robustness makes topologically ordered phases especially attractive for quantum-information applications, especially in the context of topological quantum memory and fault-tolerant quantum computation \cite{kitaev2003fault,dennis2002topological,nayak2008non,raussendorf2007topological,webster2022universal,bombin2023logical}. Beyond this stability, such phases can host exotic quasiparticles, including Abelian and non-Abelian anyons, whose exchange statistics go beyond those of ordinary bosons and fermions \cite{moore1991nonabelions,nayak2008non,tantivasadakarn2023shortest,google2023non,xu2023digital,iqbal2024non,xu2024non,iqbal2025qutrit,vojta2025topological}. These excitations enable topologically protected manipulations of quantum information. Their unconventional exchange statistics and nonlocal protection have therefore driven sustained interest across condensed-matter physics, quantum simulation, and quantum information science \cite{fujimoto2008topological,sato2010non,chen2014symmetry,fidkowski2013non,satzinger2021realizing,semeghini2021probing,google2023non,xu2023digital,iqbal2024measurement,iqbal2024non,xu2024non,iqbal2025qutrit,qin2025dynamical,will2025probing,evered2025probing,gammonSmith2026simulating,shen2026simulating}.

Although the early concept of topological order was strongly motivated by the FQHE, major theoretical progress has come from exactly solvable lattice models, most notably Kitaev’s toric code \cite{dennis2002topological,kitaev2003fault,kitaev2006topological,castelnovo2007entanglement,stark2011localization,kitaev2010topological,linsel2026independent} and the Wen--plaquette model \cite{levin2003fermions,wen2003quantum,yu2008topological,you2012projective,yu2013majorana}. These models provide canonical and analytically controlled frameworks for understanding the structure of emerging topological phases. Crucially, these models also provide tractable settings for quantum simulation and implementation on programmable quantum hardware, and have therefore played a central role in connecting topological many-body theory with experimental quantum platforms \cite{peng2014experimental,luo2015quantum,zhong2016emulating,sameti2017superconducting,li2017experimental,song2018demonstration,satzinger2021realizing,semeghini2021probing,google2023non,xu2023digital,iqbal2024measurement,iqbal2024non,xu2024non,iqbal2025qutrit,will2025probing,evered2025probing,gammonSmith2026simulating,shen2026simulating}.

Following these theoretical developments, substantial progress has been made in quantum-hardware simulations of topological phases in lattice models
\cite{shen2026simulating,peng2014experimental,luo2015quantum,sameti2017superconducting,li2017experimental,satzinger2021realizing,semeghini2021probing,xu2023digital,tantivasadakarn2023hierarchy,iqbal2024measurement,iqbal2024non,xu2024non,iqbal2025qutrit,gammonSmith2026simulating,hai2023identifying,xiang2024topological,xu2024fibonacci,minev2025stringnet,shen2026laughlin,lo2026universal}, including the first realization of the Wen--plaquette model reported in Ref.~\cite{peng2014experimental} and recent demonstrations of non-equilibrium topological order and topological matter dynamics
\cite{xiang2024topological,xu2024fibonacci,will2025probing,evered2025probing,minev2025stringnet,lo2026universal}. Building on this progress, the present work has two objectives. First, we use a tractable finite system to characterize how local plaquette stabilizers and nonlocal Wilson-loop signals remain observable on real quantum hardware. This controlled setting allows us to compare directly with classical reference calculations and to determine how equilibrium and dynamical crossover signatures are modified by static disorder, accumulated circuit noise, and effective non-Hermitian imaginary fields. Second, guided by the stabilizer structure established in the small system, we extend the implementation to a larger square lattice using a projection-inspired preparation protocol on a two-dimensional processor. Together, these two components connect quantitative characterization with a hardware-compatible route toward scalable Wen--plaquette simulations.

\section{Results}
\subsection{Models and topological diagnostics}
In this section, we introduce the Wen-plaquette models and the finite-size diagnostics used to quantify their robustness on quantum hardware. Our focus is on two observables: a local plaquette stabilizer, which probes the stabilizer-sector structure, and a nonlocal Wilson loop, which probes the persistence of finite-size loop response. For quantum simulation, guided by the properties established in the small system, we then employ a scalable stabilizer-sector preparation method to extend the hardware implementation to larger square lattices and examine whether the local plaquette structure remains observable.

\subsubsection{Wen--plaquette model and its  extensions}

We consider the spin-$1/2$ Wen--plaquette model \cite{wen2003quantum,peng2014experimental}, an exactly solvable model realizing intrinsic topological order in the thermodynamic limit,
\begin{equation}\label{wen}
H_{\rm wen}=-J\sum_{i,j}F_{i,j},
\qquad
F_{i,j}=
\sigma_{i,j}^{x}\sigma_{i+1,j}^{y}
\sigma_{i+1,j+1}^{x}\sigma_{i,j+1}^{y},
\end{equation}
where $F_{i,j}$ acts on the four spins surrounding plaquette $(i,j)$ [Fig.~\ref{fig:figure1}(a)]. Disorder and other effects are introduced below. We first study a $3\times3$ open lattice of nine qubits [Fig.~\ref{fig:figure1}(b)] and later extend the protocol to a larger lattice. Using a single plaquette index,
\begin{equation}\label{wen_finite}
H_{\rm wen}^{3\times3}=-J\sum_iF_i,
\qquad
F_i=\sigma_i^x\sigma_{i+1}^y\sigma_{i+4}^x\sigma_{i+3}^y,
\end{equation}
with
\begin{equation}\label{f}
\begin{aligned}
F_{1}&=\sigma_{1}^{x}\sigma_{2}^{y}\sigma_{5}^{x}\sigma_{4}^{y},
&
F_{2}&=\sigma_{2}^{x}\sigma_{3}^{y}\sigma_{6}^{x}\sigma_{5}^{y},\\
F_{4}&=\sigma_{4}^{x}\sigma_{5}^{y}\sigma_{8}^{x}\sigma_{7}^{y},
&
F_{5}&=\sigma_{5}^{x}\sigma_{6}^{y}\sigma_{9}^{x}\sigma_{8}^{y}.
\end{aligned}
\end{equation}
Although this finite geometry does not itself realize thermodynamic topological order, it retains the stabilizer structure and nonlocal loop diagnostics required for hardware measurements. It can also be mapped with modest overhead onto a linear IBM Quantum qubit chain (see Appendix S5). Exact solvability follows from $[F_i,F_j]=0$ and $F_i^2=\mathbb{I}$, so the eigenstates are labeled by $\{F_i=\pm1\}$. For $J>0$, the ground-state manifold $\ket{\Psi_g^A}$ belongs to the $\mathbb{Z}_{2A}$ sector with $\langle F_i\rangle=+1$; for $J<0$, $\ket{\Psi_g^B}$ belongs to the $\mathbb{Z}_{2B}$ sector with $\langle F_i\rangle=-1$.

Fig.~\ref{fig:figure1}(d) illustrates the simplest local plaquette excitation above the $\mathbb{Z}_{2A}$ sector. In the reference ground-state sector, all plaquettes satisfy $F_i=+1$. A local Pauli operation acting on a given spin flips the eigenvalues of those adjacent plaquette operators with which it anticommutes. Thus, in the bulk, a single-spin Pauli perturbation may create several neighboring plaquette defects, whereas on an open boundary or corner it can flip a single adjacent plaquette. Since each plaquette term contributes $-J F_i$ to the energy, changing one plaquette eigenvalue from $+1$ to $-1$ raises the energy by $2J$. More generally, a configuration with $n_{\rm flip}$ flipped plaquettes has an excitation energy $2J n_{\rm flip}$ above the reference stabilizer sector. This defines the elementary energy cost per local plaquette defect. On the present finite lattice, this gap provides the basic energetic protection of the stabilizer structure: weak local perturbations do not readily mix the low-energy sector unless they create such plaquette defects.

To tune the system away from the stabilizer-dominated regime, we introduce a uniform transverse field \cite{peng2014experimental,yu2008topological},
\begin{equation}\label{H}
H_{X}=H^{3\times3}_{\rm wen}-g\sum_i \sigma_i^x,
\end{equation}
where $g$ is the field strength. Since the field term does not commute with the plaquette stabilizers, increasing $g$ drives the ground state toward a trivial $x$-polarized paramagnet. In the thermodynamic limit, the transition occurs at $|J|=|g|$; on the finite open lattice, it appears as a smooth crossover visible in the local plaquette stabilizer $F_1$ and the boundary Wilson loop
\begin{equation}\label{w}
W=
\sigma_{1}^{x}\sigma_{2}^{y}\sigma_{3}^{x}\sigma_{6}^{y}
\sigma_{9}^{x}\sigma_{8}^{y}\sigma_{7}^{x}\sigma_{4}^{y}.
\end{equation}
The alternating $\sigma^x$ and $\sigma^y$ operators follow the lattice boundary and are compatible with the Wen--plaquette stabilizer structure  . Further details are provided in Appendix~\ref{Method1}.

To investigate the robustness of the Wen--plaquette regime against static inhomogeneity, we introduce on-site disorder,
\begin{equation}\label{dis}
H_{\rm dis}=H^{3\times3}_{\rm wen}-g\sum_i \sigma_i^x-\sum_i h_i \sigma_i^z,
\end{equation}
where the disorder fields are independently drawn from $h_i\in[-h/2,h/2]$. These spatially varying fields break translation invariance and compete with the coherent plaquette structure of the clean Wen model. As the disorder strength increases, the plaquette and Wilson-loop diagnostics are expected to weaken and eventually become strongly suppressed. Fig.~\ref{fig:figure1}(e) summarizes the schematic finite-size response of the clean and disordered Wen--plaquette models. In the clean limit, the competition between the plaquette coupling and the transverse field produces a crossover near $J\simeq g$, which sharpens into a quantum phase transition only in the thermodynamic limit. Disorder shifts and broadens this crossover: for moderate disorder, the stabilizer- and loop-based diagnostics remain visible, whereas in the strongly disordered regime, $J,g\ll h$, the response becomes disorder-dominated and the diagnostic crossover is washed out.

We also consider the non-Hermitian extension
\begin{equation}\label{H2}
H_{\gamma}
=
H^{3\times3}_{\rm wen}
-g\sum_i\sigma_i^x
+i\gamma\sum_i\sigma_i^z,
\end{equation}
where $\gamma$ controls an imaginary field that serves as a minimal phenomenological model of loss-like nonunitary effects, see more details in Appendix.~\ref{suppphase}. We do not implement the nonunitary propagator through postselection. Instead, $H_\gamma$ is used to generate the relevant normalized right-eigenstate or time-evolved target states, which are subsequently compiled into hardware-efficient variational circuits.

\begin{figure*}
    \centering
    \includegraphics[width=0.9\linewidth]{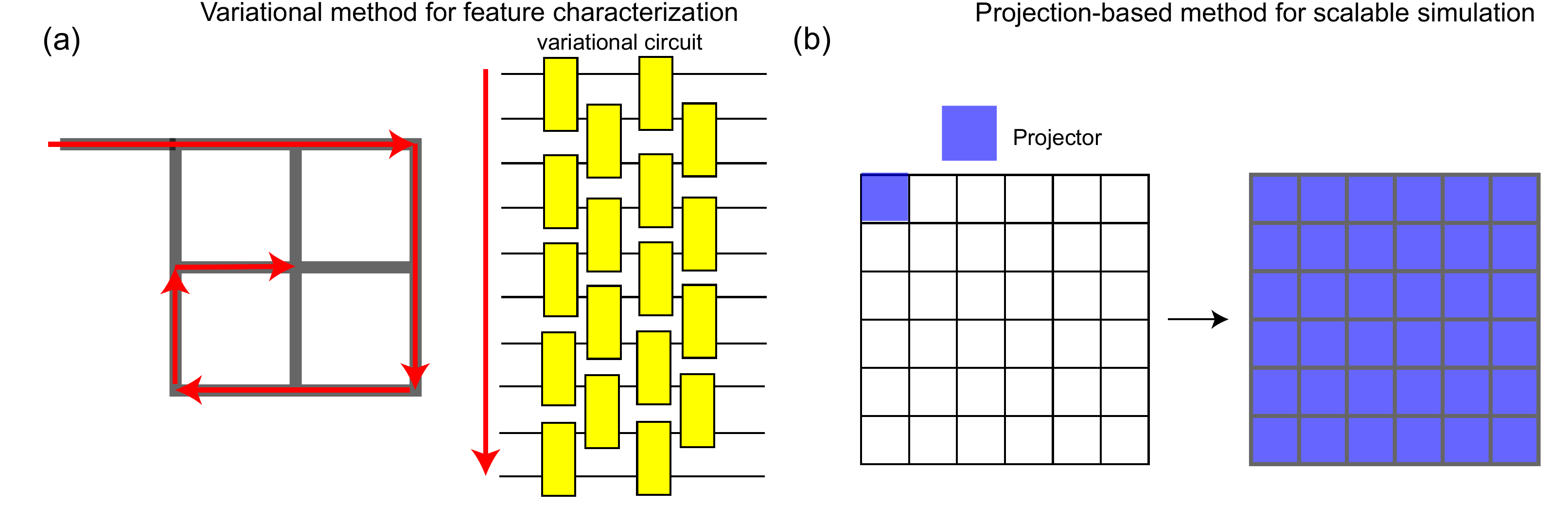}
\caption{\textbf{Simulation strategies for the Wen--plaquette model.}
\textbf{(a)} Variational method used for the detailed characterization of crossover signatures. The red arrows indicate the one-dimensional qubit ordering used to embed the two-dimensional plaquette geometry into a hardware-efficient brickwork circuit. The target states of the finite Wen--plaquette Hamiltonian [Eq.~\ref{wen_finite}] are approximated by a variational circuit composed of trainable local blocks (yellow), whose parameters are optimized using the loss function in Eq.~\ref{E}. This approach enables controlled comparisons with exact results for the small system and is used to study the effects of disorder, accumulated hardware noise, and non-Hermitian fields. Further circuit details are provided in Fig.~\ref{fig:qae} of Appendix~\ref{qae}.
\textbf{(b)} Projection-inspired method for extending the implementation to larger two-dimensional lattices. Starting from a product state, local plaquette constraints are imposed across the square lattice using the coherent representative-qubit construction motivated by the projector $\Pi_{i,j}=(\mathbb{I}+F_{i,j})/2$ [Eq.~\ref{pj}]. The resulting state lies in the target stabilizer sector, illustrated by the blue plaquettes, without requiring variational optimization or mid-circuit measurement. The corresponding large-lattice implementation is presented in Fig.~\ref{fig:large2}.
}
    \label{fig:method}
\end{figure*}

\subsection{State preparation}

Quantum circuits in this work are used to prepare the states required for measuring the plaquette stabilizer $F_1$ and the Wilson loop $W$ in both equilibrium and quench settings. We employ two complementary preparation strategies. For the general Hamiltonians in Eqs.~\ref{H}, \ref{dis}, and \ref{H2}, including their quench dynamics, we compile target states into a hardware-efficient variational circuit. For the clean commuting-projector Hamiltonian in Eq.~\ref{wen}, we additionally use its stabilizer structure to construct a representative state directly in the $\mathbb{Z}_{2A}$ ground-state sector on a larger two-dimensional processor.

For the variational workflow shown in Fig.~\ref{fig:method}(a), the target state depends on the physical task. For the Hermitian equilibrium calculations, it is the ground state of $H_X$ or $H_{\rm dis}$. For the static non-Hermitian calculations, it is the normalized right eigenstate of $H_\gamma$ whose eigenvalue has the smallest real part. For the quench calculations, the target at each time $t$ is the corresponding unitary or normalized nonunitary time-evolved state. These target states are computed classically for the nine-qubit system and then approximated by an $n$-layer circuit $V_n(\boldsymbol{\theta})$ satisfying
\begin{equation}
V_n(\boldsymbol{\theta}_{\rm opt})\ket{\phi_0}
\approx
\ket{\psi_{\rm tar}},
\end{equation}
where $\ket{\phi_0}$ is a fixed product-state input and $\ket{\psi_{\rm tar}}$ denotes the task-dependent target state. The variational parameters are obtained by minimizing
\begin{equation}\label{E}
\mathcal{C}(\boldsymbol{\theta})
=
1-\mathcal{F}\!\left[
V_n(\boldsymbol{\theta})\ket{\phi_0},
\ket{\psi_{\rm tar}}
\right],
\end{equation}
with the state fidelity $\mathcal{F}(\ket{\varphi},\ket{\psi})
=
|\bra{\varphi}\ket{\psi}|^2$. Once optimized, the compiled circuits are executed on IBM Quantum hardware, where $F_1$ and $W$ are measured to determine how the corresponding finite-size topological diagnostics are modified by hardware noise. As shown in Fig.~\ref{fig:method}, each building block consists of one ECR and two $U_{3}$ gates, and further implementation details are provided in Appendix~\ref{medv}. Unless otherwise stated, the hardware results for the $3\times3$ system are obtained using this variational compilation workflow.

For the clean Wen--plaquette Hamiltonian, the mutually commuting stabilizers also provide a direct route to the $\mathbb{Z}_{2A}$ ground-state sector. As shown in Fig.~\ref{fig:method}(b), a representative state in the simultaneous $F_{i,j}=+1$ sector can be written formally, up to normalization, as
\begin{equation}\label{pj}
\ket{G_A}
\propto
\prod_{i,j}\Pi_{i,j}\ket{0}^{\otimes N},
\qquad
\Pi_{i,j}
=
\frac{\mathbb{I}+F_{i,j}}{2}.
\end{equation}
Each projector $\Pi_{i,j}$ enforces the $+1$ eigenvalue constraint of the corresponding plaquette stabilizer, and the mutual commutativity of the $F_{i,j}$ makes the order of the formal projections irrelevant. This state is distinct from the product-state input $\ket{\phi_0}$ used in the variational workflow: $\ket{G_A}$ is itself the desired clean stabilizer-sector state. In Sec.~\ref{scale}, we implement a hardware-compatible representative-qubit construction motivated by Eq.~\ref{pj} and use it to prepare and verify the stabilizer sector on a larger square-lattice IBM quantum processor; see also Fig.~\ref{fig:large2} below.

\begin{figure}   
	\centering
	\includegraphics[width=0.99\linewidth]{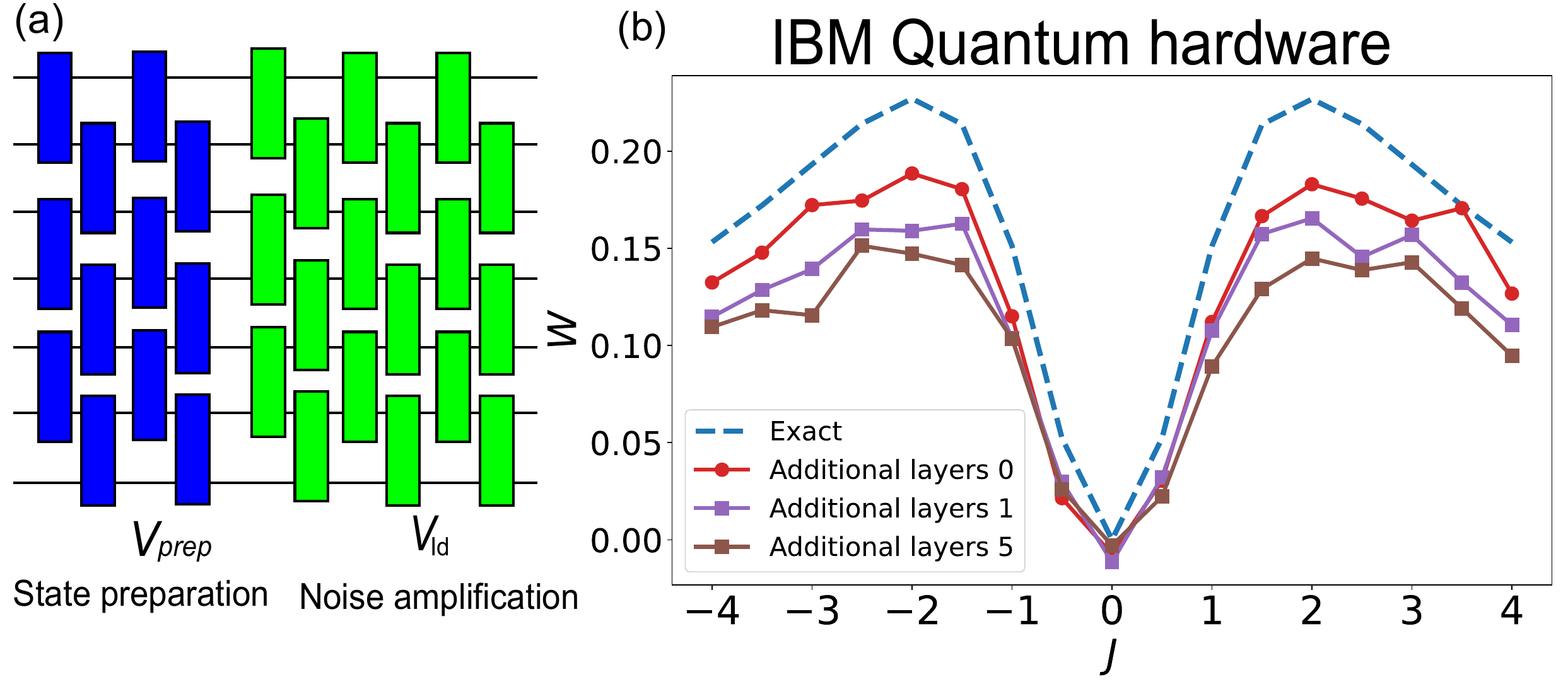}
\caption{\textbf{Noise resilience of topological diagnostics via the Wilson loop.}
\textbf{(a)} Noise-amplification protocol [Eq.~\ref{ex}]. The circuit $V_{\rm prep}$ (blue) denotes the optimized variational ansatz, while $V_{\rm Id}$ (green) denotes the inserted identity block. This construction increases the physical circuit depth, and hence the accumulated gate and decoherence errors, while leaving the ideal unitary evolution unchanged. Explicit circuit layouts and transpilation details are provided in Appendix S5.
\textbf{(b)} Robustness of the Wilson-loop crossover (between $J/g\in[1,2]$) under noise, measured on IBM Quantum hardware by preparing approximate ground states of Eq.~\eqref{H} and evaluating the large Wilson loop $W$ [Eq.~\eqref{w}]. Noise amplification is implemented by inserting $j=0,1,5$ additional identity blocks (red, violet, and brown curves, respectively). Despite the increased circuit depth, the crossover profile near $J/g\in[1,2]$ remains clearly resolved. Data are shown for the disorder-free Wen--plaquette model [Eq.~\ref{H}] at fixed $g=1$ and $J\in[-4,4]$.
}
	\label{fig:noise}
\end{figure}
    
\begin{figure*}
	\centering
	\includegraphics[width=0.7\linewidth]{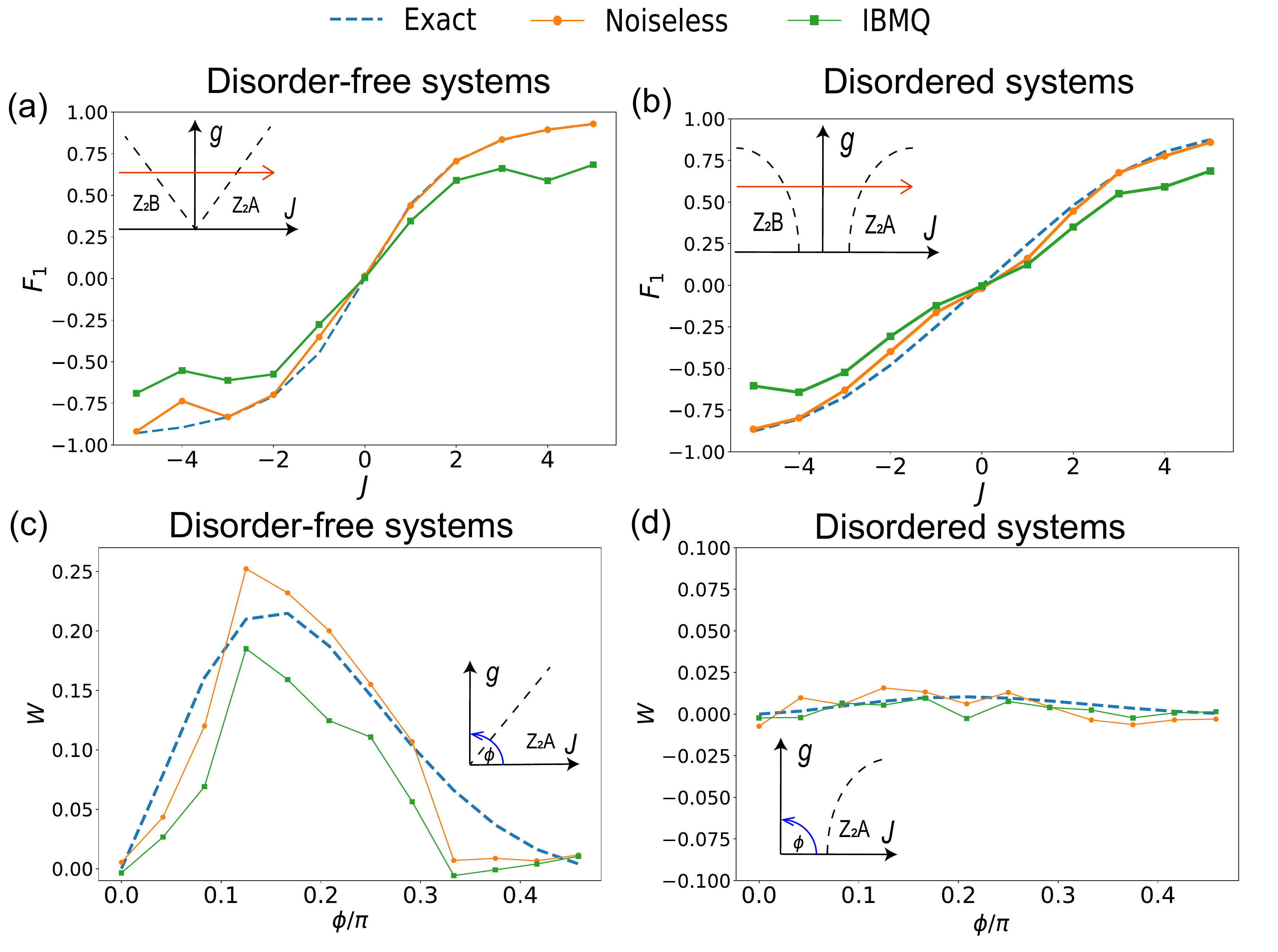}
\caption{\textbf{Measurement of finite-size stabilizer and loop diagnostics in clean and disordered Wen--plaquette models on IBM Quantum.}
We compare the disorder-free and disordered Hamiltonians in Eqs.~\eqref{H} and~\eqref{dis}, following the parameter trajectories illustrated in Fig.~\ref{fig:figure1}(e). Disorder is included in panels \textbf{(b)} and \textbf{(d)} with $h=5$, while panels \textbf{(a)} and \textbf{(c)} show the clean case, $h=0$. \textbf{(a,b)} Expectation value of the single-plaquette stabilizer $F_1$ as $J$ is varied at fixed $g=1$, revealing the finite-size crossover between the $\langle F_1\rangle\simeq-1$ and $\langle F_1\rangle\simeq+1$ stabilizer sectors. \textbf{(c,d)} Expectation value of the boundary Wilson loop $W$ [Eq.~\eqref{w}] along the path $(J,g)=(\cos\phi,\sin\phi)$, which probes the crossover between the stabilizer-dominated regime with a strong loop response and the transverse-field-dominated trivial regime. Blue, yellow, and green curves denote exact diagonalization, noiseless circuit simulations, and IBM Quantum measurements, respectively. For the Wilson-loop measurements, an ancilla-assisted interferometric protocol maps $\langle W\rangle$ onto a single ancilla-qubit population, reducing sensitivity to multi-qubit readout errors; see Fig.~\ref{fig:qae} in the Appendix. The local stabilizer-sector crossover remains clearly resolved even at $h=5$, whereas strong disorder strongly suppresses the nonlocal Wilson-loop response, demonstrating the contrasting robustness of local and nonlocal finite-size topological diagnostics. All hardware states are prepared using the variational circuit shown in Fig.~\ref{fig:method}(a).
}
	\label{fig:figure2}
\end{figure*}

\subsection{Hardware-resolved topological crossovers through variational simulation}
\subsubsection{Noise resilience of preparation and readout}
We begin with the central practical question of this work: over what range of accumulated hardware noise do finite-size topological diagnostics remain experimentally distinguishable from trivial behavior \cite{cai2023quantum,temme2017error,endo2018practical,takagi2022fundamental,knill2005quantum,preskill2018quantum,shen2025circuit}? Using the classically tractable small system as a calibrated reference, we increase the physical circuit depth by inserting identity-like variational layers that leave the ideal target state unchanged. To effectively amplify hardware noise without changing the ideal computation, we append $j$ identity layers to the optimized state-preparation circuit,
\begin{equation}\label{ex}
V_{\mathrm{extend}}(j)
=
\left(V_{\mathrm{id}}\right)^{j}V_{\mathrm{prep}},
\end{equation}
where $V_{\mathrm{prep}}$ denotes the optimized variational circuit used to prepare the target ground state (see Appendix.~\ref{medv}), and $V_{\mathrm{id}}$ is a single layer with the same gate structure as the variational ansatz but parameterized so that its noiseless action is the identity. The exponent $j$ therefore explicitly denotes the number of appended identity layers. Increasing $j$ raises the physical gate count and circuit duration, thereby increasing the accumulated gate and decoherence errors, while leaving the ideal output state unchanged: $\left(V_{\mathrm{id}}\right)^{j}
V_{\mathrm{prep}}\ket{\psi_0}
=
V_{\mathrm{prep}}\ket{\psi_0}$
in the noiseless limit. Here, $\ket{\psi_0}$ is the reference product state, and $V_{\mathrm{prep}}\ket{\psi_0}$ approximates the target ground state. As illustrated in Fig.~\ref{fig:noise}(a), varying $j$ therefore provides a controlled measure of the sensitivity of the prepared state and its Wilson-loop response to accumulated hardware noise. We apply this protocol to the disorder-free Wen--plaquette model [Eq.~\ref{H}], so that the observed degradation can be attributed primarily to hardware noise rather than to intrinsic disorder.

Fig.~\ref{fig:noise}(b) presents a hardware demonstration of the robustness of the Wilson-loop signal under controllably amplified circuit noise. We probe the finite-size crossover between the Wen--plaquette and trivial polarized regimes by sweeping $J/g$ at fixed $g=1$ and measuring the boundary Wilson loop $\langle W\rangle$ [Eq.~\ref{w}]. To read out this long Pauli string, we use an ancilla-assisted interferometric protocol: the ancilla is prepared in $\ket{+}$, a controlled-$W$ operation is applied, and a final Hadamard rotation maps the Wilson-loop expectation value onto the ancilla population,
\begin{equation}
\langle W\rangle = 2P_a(0)-1.
\end{equation}
Because the signal is extracted from a single ancilla measurement rather than from the parity of all system-qubit outcomes, this protocol reduces sensitivity to accumulated multi-qubit readout errors, although it introduces additional controlled-gate overhead (see Appendix~\ref{qae}). Two trends are evident. First, at the baseline depth ($j=0$), the measured Wilson-loop response closely follows the noiseless result, showing that the nonlocal diagnostic is preserved by the hardware implementation. Second, $\langle W\rangle$ increases rapidly for $J/g\in[1,2]$, marking the finite-size crossover into a regime with strong loop response. Importantly, this crossover remains clearly resolved as additional identity layers are inserted, demonstrating that the circuit and readout protocol retain the relevant topological signal under systematically and effectively increased hardware noise.

\subsubsection{Equilibrium diagnostics via plaquettes and Wilson loops}

Following the above analysis, we test whether the stabilizer and loop diagnostics remain distinguishable from topologically trivial behavior on quantum hardware in both clean and disordered settings. Using variationally prepared states [Fig.~\ref{fig:method}(a)], we measure a representative plaquette stabilizer and the largest boundary Wilson loop for the disorder-free Hamiltonian [Eq.~\ref{H}] and its disordered extension [Eq.~\ref{dis}], as summarized in Fig.~\ref{fig:figure2}.

Each plaquette expectation value probes the same local stabilizer structure of the Wen--plaquette model, so we use $F_{1}$ as a representative observable throughout. After preparing the variational state $V_n\ket{\psi_{0}}$, we evaluate $\langle F_{1}\rangle$ as a function of the plaquette coupling $J$ at fixed $g=1$, following the trajectory indicated in Fig.~\ref{fig:figure1}(e). For both the clean model [Fig.~\ref{fig:figure2}(a)] and the disordered model [Fig.~\ref{fig:figure2}(b)], the hardware results reproduce the same qualitative crossover structure. At sufficiently large $|J|$, the system enters stabilizer-dominated regimes in which $\langle F_{1}\rangle$ approaches $\pm1$, corresponding on the finite lattice to the two plaquette sectors selected by the sign of $J$. The persistence of this crossover profile in the presence of disorder shows that the local plaquette stabilizer remains a robust diagnostic of the finite-size Wen--plaquette structure.

To probe the weak-coupling region more directly, we sweep along the trajectory
\begin{equation}\label{phi}
J(\phi)=\cos\phi,
\qquad
g(\phi)=\sin\phi,
\end{equation}
indicated by the blue arrow in Fig.~\ref{fig:figure1}(e). Along this path, the Wilson loop $\langle W\rangle$ provides a sensitive nonlocal diagnostic. In the clean system [Fig.~\ref{fig:figure2}(c)], $\langle W\rangle$ exhibits a pronounced crossover near $\phi\simeq0.25\pi$, where $J=g$, separating the plaquette-dominated regime with a strong loop response from the transverse-field-dominated trivial regime. In the strongly disordered system [Fig.~\ref{fig:figure2}(d)], by contrast, $\langle W\rangle$ remains strongly suppressed throughout the sweep, indicating that disorder destroys the coherent nonlocal loop response. These measurements reproduce the qualitative finite-size structure obtained from the classical calculations summarized in Fig.~\ref{fig:figure1}(e) and Appendix~S4.

\subsubsection{Unitary quenches: dynamical stability across topological sectors}

\begin{figure*}
    \centering
    \includegraphics[width=0.9\linewidth]{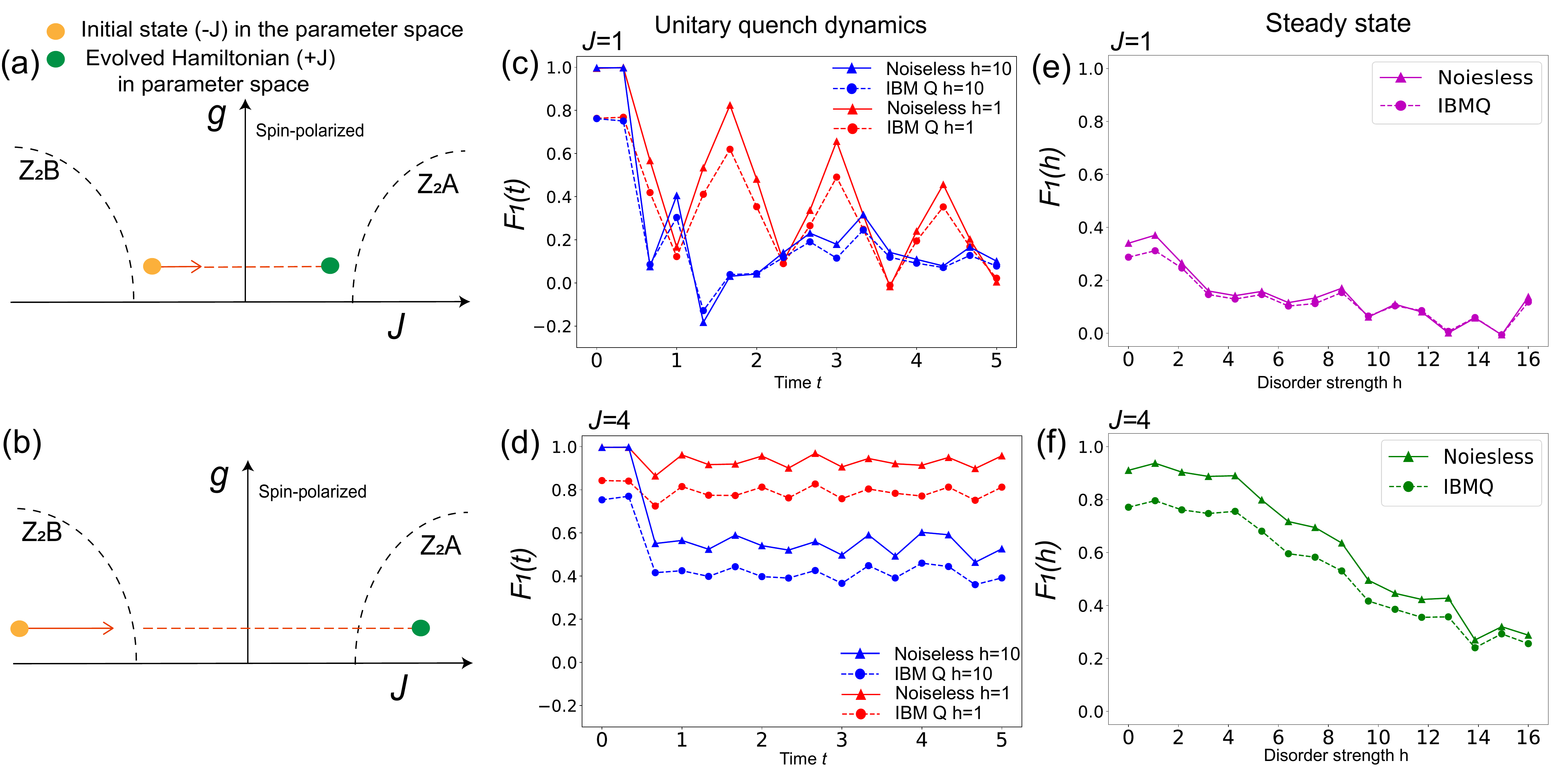}
 \caption{
\textbf{Unitary quench dynamics of finite-size plaquette diagnostics.}
\textbf{(a,b)} Finite-size response maps of the disordered model [Eq.~\ref{dis}]. Black curves denote the disorder-reshaped crossover lines obtained from exact diagonalization. Orange arrows indicate the quench protocols of Eq.~\ref{dy}: panel \textbf{(a)} considers a weak-coupling quench near the finite-size crossover, while panel \textbf{(b)} considers a strong-coupling quench deep in the stabilizer-dominated regime. The dashed orange segment connects the late-time trajectory to the post-quench Hamiltonian at $+J$ (blue dot). 
\textbf{(c,d)} Time evolution of the plaquette expectation $F_1(t)$ for initial states prepared at the red dots in panels \textbf{(a,b)}. Solid curves show noiseless results, and dashed curves show IBM Quantum measurements. We set $J=1$ in \textbf{(c)} and $J=4$ in \textbf{(d)}, and compare weak ($h=1$, red) and strong ($h=10$, blue) disorder. Near the crossover, disorder rapidly suppresses the plaquette response, whereas deep in the strong-coupling regime the response decays more slowly and remains appreciably nonzero throughout the evolution window. 
\textbf{(e,f)} Long-time response $F_1(t{=}20)$ [Eq.~\ref{fh}] as a function of disorder strength $h$ for $J=1$ (violet) and $J=4$ (green). At large disorder, the weak-coupling response becomes nearly vanishing, while the strong-coupling case retains a finite plaquette signal. The central takeaway is that the dynamical survival of the finite-size stabilizer signature is controlled by the competition between disorder and the plaquette energy scale: it is fragile near the crossover but substantially more robust in the strong-coupling regime. In all panels, $g=1$. Each dynamical snapshot is variationally compiled using the circuit architecture shown in Fig.~\ref{fig:method}(a).
}
    \label{fig:dy}
\end{figure*}

In the previous section, we characterized the equilibrium structure of the Wen--plaquette model using local plaquette expectations and nonlocal Wilson-loop observables. Because the system studied here is small, the corresponding equilibrium changes appear as smooth crossovers rather than sharp singularities. Dynamical probes nevertheless provide a useful complementary route for distinguishing different finite-size regimes on digital quantum hardware.

Here, we show that the underlying topological-to-trivial transition can also leave dynamical signatures through non-equilibrium critical-like behavior, which can be revealed using quantum quench protocols \cite{PhysRevX.11.031062}. In such a protocol, the system is initially prepared in the ground state of one Hamiltonian and then evolved under a different Hamiltonian. Even in finite systems, the resulting non-equilibrium dynamics can exhibit characteristic differences between distinct parameter regimes, thereby providing a complementary method for probing the response of the underlying phase transition.

To probe the dynamical signatures of the finite-size Wen--plaquette regimes, we simulate the following quench protocol for the unitary case:
\begin{equation}\label{dy}
\ket{\psi(t)}=e^{-itH_{\rm dis}}\ket{\psi_{-J}},
\end{equation}
where $\ket{\psi_{-J}}$ denotes the ground state of the disorder-free Hamiltonian [Eq.~\ref{H}] with coupling $-J$, and $H_{\rm dis}$ is the disordered Hamiltonian with coupling $+J$. This procedure effectively drives the system between the two stabilizer sectors distinguished by the sign of $J$. By preparing the system in one stabilizer-dominated regime and suddenly quenching to the opposite parameter regime, we can monitor how the state evolves under unitary dynamics and identify dynamical signatures associated with the change between these distinct finite-size configurations.

For each evolution time $t$, the state $\ket{\psi(t)}$ is used as the target for variational compilation, yielding a hardware-efficient circuit $V_n(t)$ such that
\begin{equation}
V_n(t)\ket{\psi_0}\approx\ket{\psi(t)}
\end{equation}
(see Sec.~\ref{medv} in Methods for details). We then measure the plaquette stabilizer $F_1$ on the variationally prepared state. The quench reverses the sign of the plaquette coupling, from an initial ground state prepared at $-J$ to evolution under a disordered Hamiltonian with coupling $+J$. It therefore switches the energetically favored plaquette sector from $\mathbb{Z}_{2B}$ to $\mathbb{Z}_{2A}$. We compare two realizations of this sector-changing protocol, illustrated in Figs.~\ref{fig:dy}(a) and \ref{fig:dy}(b). For $J=1$, the plaquette and transverse-field energy scales are comparable, and both the initial and post-quench Hamiltonians lie near the finite-size topological crossover. For $J=4$, the initial and post-quench Hamiltonians lie deep in the two opposite stabilizer-dominated sectors. The purpose of this comparison is therefore not to realize a direct topological-to-trivial quench, but to determine how the dynamical stability of the plaquette-sector signature depends on the distance from the finite-size crossover.

Near the crossover at $J=1$, disorder competes efficiently with the relatively weak plaquette energy scale. Consequently, the plaquette response rapidly decreases toward small values, particularly for strong disorder, as shown by the blue curve in Fig.~\ref{fig:dy}(c). By contrast, for $J=4$, the larger energy cost of creating plaquette defects suppresses disorder-induced mixing between stabilizer sectors. The responses in Fig.~\ref{fig:dy}(d) therefore decay more slowly and remain appreciably nonzero throughout the simulated time window. The blue curve still exhibits a visible reduction, but this reduction is substantially weaker than in the near-crossover case. These results show that the dynamical survival of the finite-size topological diagnostic is controlled by the competition between the disorder strength and the stabilizer energy scale, rather than simply by the presence of a nonzero initial plaquette signal. We also observe an approximately time-independent offset between the noiseless and hardware results in Fig.~\ref{fig:dy}(d). The persistence of this offset over time suggests a systematic attenuation of the measured plaquette expectation, arising from accumulated gate and readout errors, rather than an additional hardware-induced decay of the dynamics. Accordingly, the hardware data reproduce the temporal stability of the plaquette response more accurately than its absolute magnitude. The results shown here are reported without correcting this systematic offset; in principle, it could be reduced through calibrated readout mitigation or reference-based rescaling.

To further quantify the disorder dependence, we evaluate the long-time plaquette response
\begin{equation}\label{fh}
F(h)=\bra{\psi(t)}F_{1}\ket{\psi(t)}\big|_{t=20},
\end{equation}
with the post-quench state
\begin{equation}
\ket{\psi(t)}=e^{-itH_{\rm dis}(h)}\ket{\psi_{-J}}.
\end{equation}
As shown in Figs.~\ref{fig:dy}(e) and \ref{fig:dy}(f), the weak-coupling regime ($J=1$) rapidly loses its plaquette signal as the disorder strength increases. By contrast, in the strong-coupling regime, the long-time response remains nonzero over a broad range of $h$. A pronounced reduction appears only when the disorder strength becomes comparable to the stabilizer energy scale, $4J=16$ [Fig.~\ref{fig:figure1}(d)]. These results demonstrate that the finite-size stabilizer structure acquires substantially greater dynamical robustness upon entering the strong-coupling regime, whereas it remains fragile near the weak-coupling crossover.

\begin{figure*}
    \centering
    \includegraphics[width=0.8\linewidth]{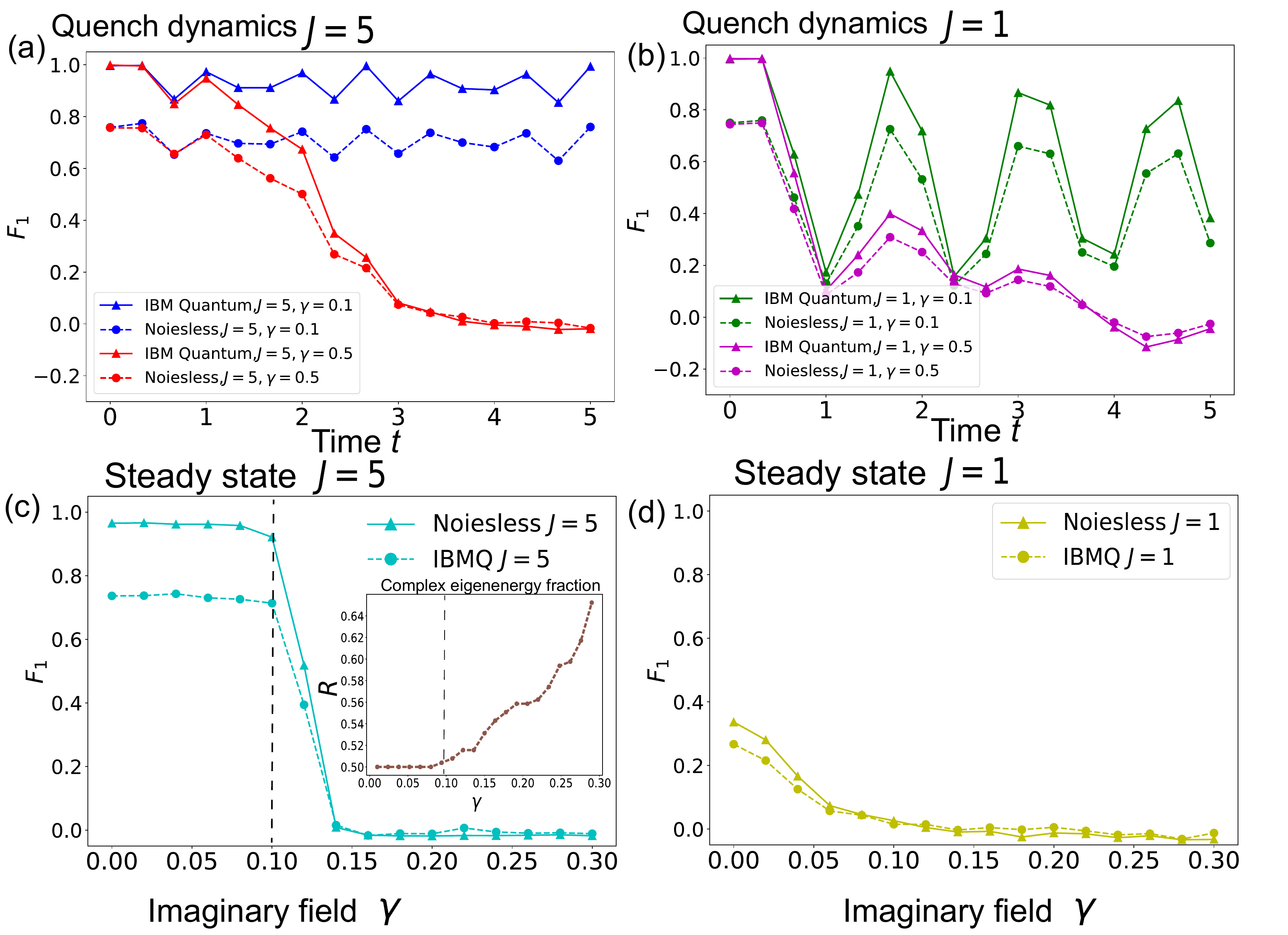}
\caption{{\bf Signatures of finite-size stabilizer response under non-Hermitian quenches.}
We study the effect of the non-Hermitian model defined by Eq.~\ref{H2}. Here, we simulate the normalized right-state evolution in Eq.~\ref{non} and monitor the plaquette expectation $\langle F_{1}\rangle$. 
\textbf{(a,b)} Time-dependent $\langle F_{1}(t)\rangle$ for strong coupling ($J=5$) and weak coupling ($J=1$), respectively. Solid lines correspond to noiseless circuit simulations, and dashed lines to IBM Quantum measurements. We compare weak ($\gamma=0.1$, blue and green) with stronger fields ($\gamma=0.5$, red and violet). For $\gamma=0.1$, the strong-coupling case preserves a large plaquette response over the simulated window, whereas larger $\gamma$ progressively suppresses the signal. 
\textbf{(c,d)} Long-time value $\langle F_{1}(t{=}20)\rangle$ [Eq.~\ref{longtime}] versus $\gamma$ for $J=5$ in (c) and $J=1$ in (d). In the strong-coupling case, $\langle F_{1}\rangle$ remains large for weak perturbation and then decreases across a crossover near $\gamma\simeq 0.1$. This crossover coincides with the emergence of a nonzero fraction of eigenvalues with finite imaginary parts, quantified by $R=\mathrm{Num}(|\mathrm{Im}\,E|>0)/\mathrm{Num}_{\rm total}$ (inset), signaling the onset of appreciable nonunitary spectral broadening. In contrast, for weak coupling, even modest $\gamma$ strongly reduces the long-time plaquette response. Across all panels, the close agreement between noiseless (solid) and experimental (dashed) curves indicates that the implemented protocol faithfully reproduces the target normalized right-state dynamics. The transverse-field strength is fixed to $g=1$. All simulations are conducted on the variational circuit [Fig.~\ref{fig:method}(a)].
}
    \label{fig:dis}
\end{figure*}

\subsubsection{Simulation of nonunitary quenches}

We next extend the above quench protocol to the non-Hermitian regime. For each observation time $t$, we first evolve the initial state under the full non-Hermitian propagator and then normalize the resulting right state for state preparation and observable evaluation:
\begin{equation}\label{non}
\ket{\psi^{\prime}(t)}
=
\frac{e^{-itH_{\gamma}}\ket{\psi_{-J}}}
{\left\|e^{-itH_{\gamma}}\ket{\psi_{-J}}\right\|}.
\end{equation}
The normalization in Eq.~\ref{non} is performed only after the evolution to the selected time $t$; no intermediate normalization is applied during the nonunitary dynamics. Accordingly, all observables in this subsection are evaluated as normalized right-state expectation values. Here, $\ket{\psi_{-J}}$ denotes the ground state of the clean Hermitian model ($\gamma=0$ and $h=0$) at plaquette coupling $-J$. After the quench, the state evolves under the non-Hermitian Hamiltonian $H_{\gamma}$ with nonzero $\gamma$ and coupling $+J$. Since the nonunitary propagator is not implemented directly on the gate-based hardware, the normalized state $\ket{\psi^{\prime}(t)}$ at each selected time is used as the target for variational compilation, $V_n(t)\ket{\psi_0}\approx\ket{\psi^{\prime}(t)},$
as detailed in Sec.~\ref{medv} of Methods.

As shown in Fig.~\ref{fig:dis}(a), we first consider the strong-coupling case at $J=5$, where the initial state is prepared deep inside the stabilizer-dominated regime. In this parameter range, the time-dependent plaquette response $\langle F_{1}(t)\rangle$ remains large under weak non-Hermitian effects, $\gamma=0.1$ for the blue curves, throughout the full evolution window shown. This behavior indicates that the finite-size stabilizer structure survives weak nonunitary effects if the coherent plaquette scale set by $J$ is large. When $\gamma$ is increased to $0.5$,  the responses represented by the red curves decay much more visibly, showing that stronger nonunitary evolution progressively breaks the plaquette-sector signal. Nevertheless, the early-time response remains at a high level, which indicates that the loss of stabilizer structure is not instantaneous but develops over evolution.

Then, in Fig.~\ref{fig:dis}(b), we show that the weak-coupling case $J=1$ is substantially more fragile. At fixed $g=1$, this parameter choice lies close to the finite-size crossover, where the plaquette interaction and transverse field compete on comparable energy scales and the stabilizer polarization is already weakened. Moreover, the imaginary field $i\gamma\sum_i\sigma_i^z$ anticommutes with the $\sigma^x$ and $\sigma^y$ factors entering the adjacent plaquette operators, and therefore promotes the creation and mixing of plaquette defects. Consequently, even $\gamma=0.1$ causes $\langle F_1(t)\rangle$ to decay rapidly, while $\gamma=0.5$ suppresses the response more strongly. By contrast, for $J=5$, the system lies deep in the stabilizer-dominated Wen--plaquette regime, where the larger defect energy scale set by $J$ suppresses such mixing. The comparison therefore provides a dynamical signature of the finite-size topological crossover: the plaquette response is robust far inside the Wen--plaquette regime but becomes increasingly fragile as the system approaches the crossover to the trivial polarized regime.

To quantify this distinction, we evaluate the finite-time response
\begin{equation}\label{longtime}
F(\gamma)
=
\bra{\psi^{\prime}(\gamma,t)}
F_1
\ket{\psi^{\prime}(\gamma,t)}
\big|_{t=20}.
\end{equation}
In the strong-coupling regime [Fig.~\ref{fig:dis}(c)], $F(\gamma)$ remains close to its large-$J$ value for weak $\gamma$, reflecting the energetic protection of the stabilizer sector against non-Hermitian defect generation. It then decreases across a finite-size crossover near $\gamma\simeq0.1$, beyond which the imaginary-field scale becomes sufficiently large to substantially mix the plaquette sectors. The inset shows that this suppression occurs in the same parameter range in which a nonzero fraction of the spectrum becomes complex,
\begin{equation}
R=
\frac{\mathrm{Num}\!\left(|\mathrm{Im}\,E|>0\right)}
{\mathrm{Num}_{\rm total}}.
\end{equation}
The simultaneous growth of $R$ and reduction of $F(\gamma)$ indicate that the loss of the plaquette response is correlated with the onset of appreciable non-Hermitian spectral complex fraction. In the weak-coupling regime [Fig.~\ref{fig:dis}(d)], no comparable plateau is observed. Because $J$ is already comparable to $g$, the system lies near the finite-size topological crossover and the stabilizer structure is only weakly protected. A modest imaginary field is therefore sufficient to mix plaquette sectors and strongly suppress the long-time response. The contrast between Figs.~\ref{fig:dis}(c) and \ref{fig:dis}(d) shows that the nonunitary dynamics retain a clear memory of the underlying equilibrium crossover: states deep in the Wen--plaquette regime exhibit a robust plaquette response, whereas states near the crossover are rapidly destabilized.

\begin{figure*}
    \centering
    \includegraphics[width=0.9\linewidth]{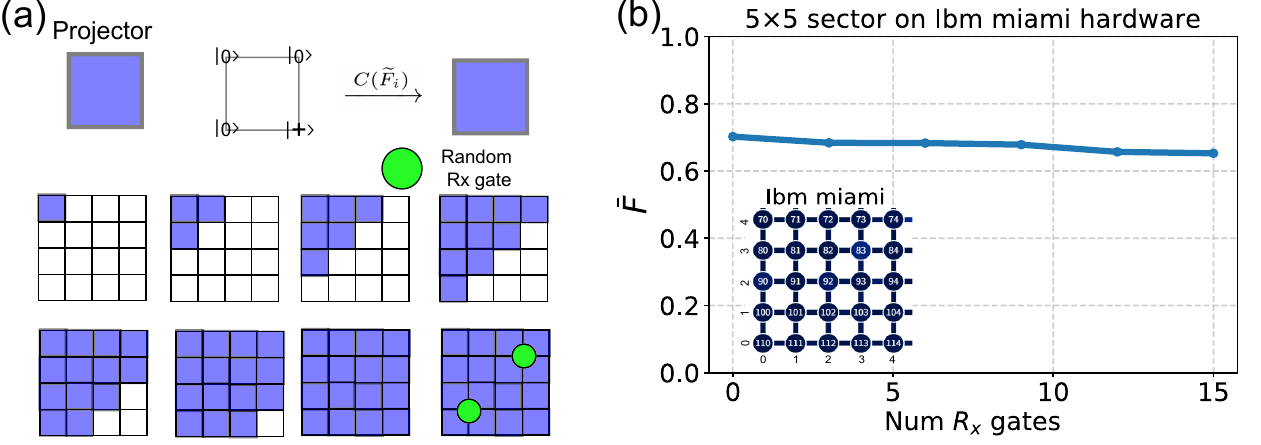}
    \caption{Square-lattice {non-variational} stabilizer-sector preparation and robustness benchmark.  (a) Schematic of the coherent representative-qubit construction used to generate local Wen--plaquette stabilizer-sector structure on a square-lattice layout.  
Each plaquette (blue) is projected onto the $+1$ stabilizer subspace using the local projector [Eq.~\ref{cf}], which is applied across the entire lattice. After completing the full projection, sparse single-qubit $R_x$ rotations (green circles) are inserted to emulate controlled local perturbations. This protocol prepares a larger-scale topological state without variational optimization, and implementation details are provided in Appendix S6. (b) Hardware demonstration on the 2D IBM miami device. We select a $5\times5$ sector (layout shown in (a)), and evaluate the average plaquette stabilizer $\bar{F}$ over the entire lattice as a function of the number of inserted coherent {random} single-qubit $R_x$ perturbations with rotation angles drawn from $[-0.1,0.1]$. The insignificant decay indicates robustness against imperfections. Data in (b) are averaged over five random realizations.}
    \label{fig:large2}
\end{figure*}

\subsection{Scalable stabilizer-projection simulations for larger lattices}\label{scale}

Here, we further discuss how a large-scale Wen--plaquette stabilizer-sector state can be prepared without resorting to variational training. As shown in Fig.~\ref{fig:method}(b), the clean Wen--plaquette model admits a direct projection-based preparation scheme in which the stabilizer constraints are sequentially enforced. Starting from a trivial product state, a representative state in the target sector can be written as $\prod_{i,j}\Pi_{i,j}\ket{0}^{\otimes N}$ [Eq.~\ref{pj}], where each local projector is $\Pi_{i,j}=\frac{\mathbb{I}+F_{i,j}}{2}.$
Acting with $\Pi_{i,j}$ projects the system onto the $+1$ eigenspace of the corresponding plaquette operator. Since all Wen--plaquette stabilizers mutually commute, these projectors may be applied in any order, greatly simplifying the preparation. A schematic overview of the protocol is shown in Fig.~\ref{fig:large2}(a). The left panel illustrates the sweeping pattern used to apply the projectors across the lattice, thereby building up the target stabilizer-sector state. The right panel depicts an additional perturbative unitary step, in which sparse random single-qubit rotations are applied throughout the system.

For the hardware demonstration, we adopt a coherent representative-qubit construction \cite{xu2023digital}, which generates the desired local stabilizer-sector structure for a prescribed plaquette input configuration without requiring mid-circuit measurement. For each plaquette stabilizer $F_{i,j}$, we choose one corner qubit $r$ as a representative control qubit and coherently generate a map proportional to $(\mathbb{I}\pm F_{i,j})$ using a controlled Pauli string on the remaining corners. Writing
\begin{equation}
F_{i,j}=P_r\,\widetilde F_{i,j},\qquad P_r\in\{\sigma_r^x,\sigma_r^y\},
\end{equation}
where $P_r$ acts on the representative qubit $r$ and $\widetilde F_{i,j}$ is the product of the remaining Pauli factors on the plaquette, we define
\begin{equation}
C_r(\widetilde F_{i,j})
=
\ket{0}\!\bra{0}_r\otimes \mathbb{I}
+
\ket{1}\!\bra{1}_r\otimes \widetilde F_{i,j}.
\end{equation}
If $P_r=\sigma_r^x$, preparing $r$ in $\ket{+_{x}}_r=(\ket{0}_r+\ket{1}_r)/\sqrt{2}$ gives
\begin{equation}\label{cf}
\begin{aligned}
C_r(\widetilde F_{i,j}) \ket{+_{x}}_r\ket{000}
&=
\frac{1}{\sqrt{2}}
\Big(
\ket{0}_r\ket{000}
+
\ket{1}_r\,\widetilde F_{i,j}\ket{000}
\Big)\\
&=
\frac{1}{\sqrt{2}}(\mathbb{I}+F_{i,j})\ket{0}_r\ket{000},
\end{aligned}
\end{equation}
where we use $\ket{1}_r=\sigma_r^x\ket{0}_r$. An analogous construction holds for $P_r=\sigma_r^y$, with the representative qubit prepared in the state $\ket{+_y}_r=(\ket{0}+i\ket{1})/\sqrt{2}$. More details are provided in Appendix S6.

In our implementation, we choose the representative qubit to be the upper-right corner, $r=(i+1,j+1)$. For the plaquette convention used here, this gives $F_{i,j}=\sigma^{x/y}_{i+1,j+1}\,\widetilde F_{i,j},$
so the corresponding input state for the coherent projector construction is $\ket{+_{x/y}}_{i+1,j+1}\ket{0}_{i,j}\ket{0}_{i+1,j}\ket{0}_{i,j+1}$ [Fig.~\ref{fig:large2} (a)].

We next implement this stabilizer-sector preparation protocol on IBM Quantum hardware. Taking advantage of the native two-dimensional connectivity of the IBM \texttt{ibm\_miami} device, we prepare a $5\times5$ state in the $\mathbb{Z}_{2A}$ stabilizer sector and measure the lattice-averaged plaquette response,
\begin{equation}
\bar{F}
=
\frac{1}{\mathrm{Num}(F)}
\sum_{i,j}
\langle F_{i,j}\rangle,
\end{equation}
where $\mathrm{Num}(F)$ is the number of measured plaquettes. This prepared state represents the stabilizer-dominated endpoint of the finite-size crossover studied in the smaller variational simulations; the corresponding trivial endpoint at large transverse field is the $\sigma^x$-polarized product state.

To test the local stability of the prepared sector, we apply additional single-qubit $R_x(\theta)$ rotations to randomly selected qubits after state preparation [Fig.~\ref{fig:large2}(a)]. The rotation angles are drawn from a narrow interval around zero, but their magnitudes are deliberately chosen to exceed the typical residual coherent error associated with a calibrated native single-qubit gate. Each inserted rotation therefore represents a controlled, intentionally amplified local perturbation rather than merely the intrinsic error of an additional hardware gate. As shown in Fig.~\ref{fig:large2}(b), $\bar F$ decreases only weakly over the investigated range of up to 15 inserted rotations and remains close to its unperturbed value. Thus, the prepared stabilizer response remains resolvable even under local coherent perturbations stronger than the nominal single-qubit control-error level of the device.

Moreover, as illustrated in Fig.~\ref{fig:large2}(a), the full preparation circuit retains a layered local structure: plaquette operations acting on non-overlapping regions can be executed in parallel within the same layer. Consequently, increasing the lattice size primarily increases the number of repeated local layers rather than requiring a fully nonlocal circuit, which provides a favorable route toward larger-scale implementations. Scaling the protocol to substantially larger systems will thus require sufficiently low two-qubit error rates, and the detailed estimation of noise effects is provided in Appendix.~\ref{noise}. In Appendix~\ref{rep}, we also discuss alternative scalable preparation schemes, including ancilla-assisted projection, postselection, and measurement-based feedforward. Their practical efficiency depends on the capabilities of the specific hardware platform, such as the availability of mid-circuit measurement and real-time classical control.


\section{Discussion}\label{discussion}

In this work, we used programmable quantum circuits to determine how finite-size Wen--plaquette signatures survive experimentally relevant imperfections. The central result is not a thermodynamic demonstration of topological order on present-day devices, but a hardware-resolved study of the parameter and noise regimes over which finite-size Wen--plaquette features remain measurable and distinguishable from trivial behavior. Relative to prior work, which has primarily emphasized the preparation, detection, or manipulation of topological states and anyonic excitations on programmable quantum platforms \cite{peng2014experimental,satzinger2021realizing,semeghini2021probing,xu2023digital,tantivasadakarn2023hierarchy,iqbal2024measurement,iqbal2024non,xu2024non,iqbal2025qutrit,will2025probing,evered2025probing,gammonSmith2026simulating}, our study examines their robustness within a unified Wen--plaquette framework that incorporates static disorder, accumulated hardware noise, and effective non-Hermitian imaginary fields. The persistence of the topological stabilizer signature is further demonstrated through a complementary square-lattice stabilizer-sector preparation, which provides a hardware-compatible route toward larger systems beyond the minimal variational implementation.

From the perspective of near-term implementation, these results help clarify a realistic path toward larger-scale simulations. In particular, the projection-based preparation protocol shows that scalable stabilizer-sector construction need not rely exclusively on deep variational optimization, and therefore offers a complementary route for studying larger Wen--plaquette systems on hardware with suitable connectivity \cite{piroli2021quantum,lu2022measurement,xu2023digital,tantivasadakarn2023hierarchy,iqbal2024measurement,iqbal2025qutrit,evered2025probing,gammonSmith2026simulating}. Looking forward, simulations on larger devices will help determine how the finite-size robustness identified here evolves toward sharper collective behavior as the system size increases \cite{will2025probing,evered2025probing,linsel2026independent,gammonSmith2026simulating}.

\section*{ACKNOWLEDGMENTS}\label{sec}
We acknowledge the use of IBM Quantum services for this work. The views expressed are those of the authors and do not reflect the official policy or position of IBM or the IBM Quantum team. All data and code for this work are available from the corresponding authors upon reasonable request. {This work is supported by the Singapore Ministry of Education Academic Research Fund Tier-II Grant (Awards MOE-T2EP50224-0007 and MOE-T2EP50224-0021)).}

\bibliography{ref}
\onecolumngrid
\flushbottom
\newpage
\appendix
\setcounter{equation}{0}
\setcounter{figure}{0}
\setcounter{table}{0}
\setcounter{section}{0}
\renewcommand{\theequation}{S\arabic{equation}}
\renewcommand{\thefigure}{S\arabic{figure}}
\renewcommand{\thesection}{S\arabic{section}}
\renewcommand{\thepage}{S\arabic{page}}

\newpage
\section*{Appendix}

\section{Gauge-invariant construction of Wilson-loop diagnostics}\label{Method1}

The construction of the Wilson-loop diagnostic in Eq.~\ref{w} is rooted in gauge invariance. Although the Wen--plaquette model is not formulated as a conventional lattice gauge theory, an emergent $\mathbb{Z}_{2}$ gauge structure becomes explicit upon representing the spins in terms of Majorana fermions $\chi_i^\alpha$. Introducing four Majorana operators per site, the Pauli operators may be written as
\begin{equation}
\sigma_i^x = i\chi_i^{\bar y}\chi_i^x,\qquad
\sigma_i^y = i\chi_i^{\bar x}\chi_i^y,\qquad
\sigma_i^z = i\chi_i^x\chi_i^{\bar x},
\end{equation}
together with the local constraint
\begin{equation}
\chi_i^x\chi_i^y\chi_i^{\bar x}\chi_i^{\bar y}=1,
\end{equation}
which projects the enlarged Majorana Hilbert space onto the physical spin sector.

In this representation, $\mathbb{Z}_{2}$ link operators arise naturally by pairing nearest-neighbour Majoranas along the edges of the Wen--plaquette model:
\begin{equation}
\begin{aligned}
K_{i,i+\hat x} &= i\chi_i^x\chi_{i+\hat x}^{\bar x},\\
K_{i+\hat x,i+\hat x+\hat y} &= i\chi_{i+\hat x}^y\chi_{i+\hat x+\hat y}^{\bar y},\\
K_{i+\hat x+\hat y,i+\hat y} &= i\chi_{i+\hat x+\hat y}^x\chi_{i+\hat y}^{\bar x},\\
K_{i+\hat y,i} &= i\chi_i^{\bar y}\chi_{i+\hat y}^y.
\end{aligned}
\end{equation}
The plaquette operator is then given by the closed product of the four link operators around a plaquette,
\begin{equation}
F_i
=
K_{i,i+\hat x}
K_{i+\hat x,i+\hat x+\hat y}
K_{i+\hat x+\hat y,i+\hat y}
K_{i+\hat y,i}.
\end{equation}

Under a local $\mathbb{Z}_{2}$ gauge transformation at site $i$, the four Majorana operators on that site transform as
\begin{equation}
\chi_i^x,\chi_i^{\bar x},\chi_i^y,\chi_i^{\bar y}
\longrightarrow
-\chi_i^x,-\chi_i^{\bar x},-\chi_i^y,-\chi_i^{\bar y}.
\end{equation}
A single link operator attached to site $i$ therefore changes sign under this transformation. By contrast, the plaquette operator $F_i$ remains invariant because each site contributes an even number of Majorana operators to the closed product. More generally, any closed product of link operators is gauge invariant. The Wilson loop used in the main text,
\begin{equation}
W=\sigma_{1}^{x}\sigma_{2}^{y}\sigma_{3}^{x}\sigma_{6}^{y}\sigma_{9}^{x}\sigma_{8}^{y}\sigma_{7}^{x}\sigma_{4}^{y},
\end{equation}
is precisely such a closed product and therefore serves as a natural nonlocal diagnostic of the finite-size Wen--plaquette regime.

\section{Variational circuits}\label{medv}
Digital simulations reported in Figs.~\ref{fig:noise}, \ref{fig:figure2}, \ref{fig:dy}, and \ref{fig:dis} are implemented using hardware-efficient variational quantum circuits
\cite{chen2020variational,cerezo2021variational,gong2024quantum,koh2022simulation,shen2023observation,koh2025interacting,chen2023high,chen2023robust,shen2024enhanced}. For an $N$-qubit register, we employ an $n$-layer ansatz of the form
\begin{equation}\label{suppv}
V_n=\prod_{k=1}^{n} V^{(k)},
\end{equation}
where each layer $V^{(k)}$ consists of alternating entangling blocks on even and odd bonds, interleaved with single-qubit Euler rotations:
\begin{equation}\label{eq:layer_def}
\begin{aligned}
V^{(k)}=
&\Bigg[
\prod_{j\in \mathrm{even}} {\rm ECR}_{j,j+1}\,
\prod_{j=0}^{N-1} U^{3}_{k,j}\!\big(\theta_{k,j},\phi_{k,j},\lambda_{k,j}\big)
\Bigg] \\
&\times
\Bigg[
\prod_{j\in \mathrm{odd}} {\rm ECR}_{j,j+1}\,
\prod_{j=0}^{N-1} U^{3}_{k,j}\!\big(\theta'_{k,j},\phi'_{k,j},\lambda'_{k,j}\big)
\Bigg].
\end{aligned}
\end{equation}
Each $U^{3}_{k,j}$ is a single-qubit parameterized rotation acting on qubit $j$ in layer $k$, with the standard form
\begin{equation}\label{eq:u3_def}
U^{3}(\theta,\phi,\lambda)=
\begin{pmatrix}
\cos\!\left(\frac{\theta}{2}\right) & -e^{i\lambda}\sin\!\left(\frac{\theta}{2}\right)\\
e^{i\phi}\sin\!\left(\frac{\theta}{2}\right) & e^{i(\phi+\lambda)}\cos\!\left(\frac{\theta}{2}\right)
\end{pmatrix}.
\end{equation}
The two-qubit entangling gate ${\rm ECR}_{j,j+1}$ denotes the echoed cross-resonance (ECR) gate used on IBM hardware
\cite{javadi2024quantum}, represented in the computational basis as
\begin{equation}\label{eq:ecr_def}
{\rm ECR}=
\frac{1}{\sqrt{2}}
\begin{pmatrix}
0 & 0 & 1 & i\\
0 & 0 & i & 1\\
1 & -i & 0 & 0\\
-i & 1 & 0 & 0
\end{pmatrix}.
\end{equation}

To approximate a target state $\ket{\psi_{\rm target}}$, we optimize the variational parameters by minimizing the fidelity loss
\begin{equation}
\begin{aligned}
C = 1 - F\!\left(V_n\ket{\psi_0},\,\ket{\psi_{\rm target}}\right),
\end{aligned}
\end{equation}
where $F(\ket{\phi},\ket{\psi}) = |\langle\phi|\psi\rangle|^2$, $\ket{\psi_0}$ is a chosen reference state, and the parameters in $V_n$ are collectively denoted by the angles $\{\theta,\phi,\lambda\}$ across all layers and qubits. Each experimental data point is estimated from $20000$ shots on IBM Quantum devices
\cite{steffen2011quantum,santos2016ibm,ibm}, providing stable statistics in the presence of hardware noise.

\section{Ancilla-assisted interferometric readout}\label{qae}

In this work, we employ an ancilla-assisted interferometric readout protocol, equivalent to a Hadamard-test measurement, to evaluate expectation values of Pauli-string observables while reducing susceptibility to multi-qubit readout errors \cite{cleve1998quantum,ekert2002direct,crawford2021efficient}. We consider an operator $O$, such as a Pauli string, and the corresponding target state $\ket{\psi}$. The protocol introduces an ancilla qubit:
\begin{equation}
\ket{\Psi_0}
=
\frac{1}{\sqrt{2}}\big(\ket{0_a}+\ket{1_a}\big)\ket{\psi}.
\end{equation}
Applying a controlled-$O$ gate, followed by a Hadamard gate on the ancilla, yields
\begin{equation}\label{eq:qae_hadamard_test}
\ket{\Psi_1}
=
\frac{1}{2}\Big[\ket{0_a}(I+O)\ket{\psi}+\ket{1_a}(I-O)\ket{\psi}\Big].
\end{equation}
A $Z$-basis measurement of the ancilla returns outcome $0$ with probability
\begin{equation}\label{eq:qae_prob}
P(0)=\frac{1}{2}\big(1+\langle O\rangle\big),
\qquad
\langle O\rangle=\bra{\psi}O\ket{\psi},
\end{equation}
leading to
\begin{equation}
\langle O\rangle = 2P(0)-1.
\end{equation}
 In contrast to direct measurement of a long Pauli string, which requires readout of all involved system qubits and is therefore vulnerable to any single readout error among them, this protocol concentrates the relevant information into a single-ancilla measurement. 

\begin{figure}[h]
    \centering
    \includegraphics[width=0.7\linewidth]{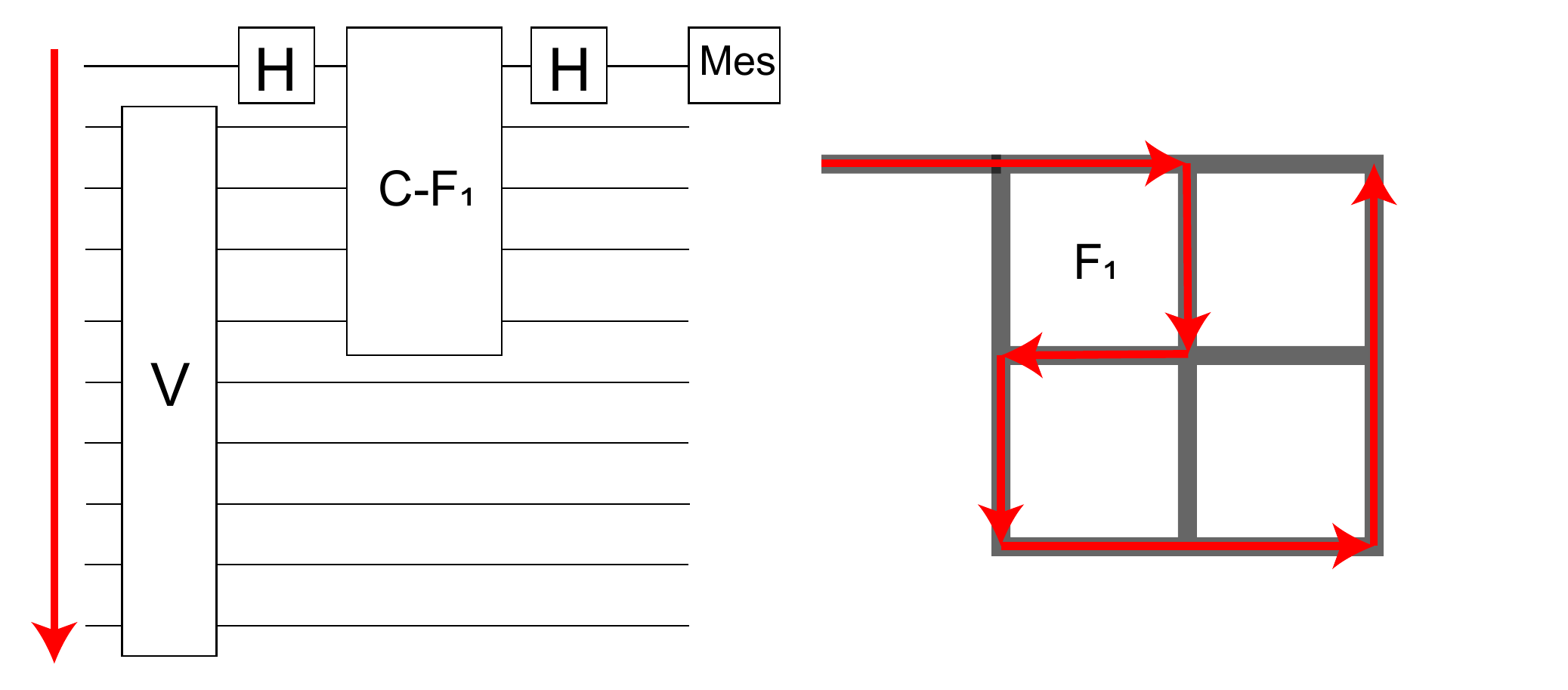}
    \caption{Workflow for the ancilla-based measurement of the plaquette operator.
The circuit first prepares the ground state using a variational circuit 
$V$. An ancilla qubit, initialized in 0, is then placed in superposition by a Hadamard gate. A controlled-$F_1$ operation entangles the ancilla with the system qubits. Measurement of the ancilla in the Z basis yields the probability $P(0)$, leading to $\langle F_{1} \rangle =2P(0)-1$.}
    \label{fig:qae}
\end{figure}

As a concrete example, to measure the four-body plaquette operator $F_{1}$, we first prepare the system register in the ground state $\ket{\psi_g}$ using a variational circuit $V$, namely $\ket{\psi_g}=V\ket{\psi_0}$. An ancilla qubit is initialized in $\ket{0_a}$, a Hadamard gate creates the required superposition, and a controlled-$F_1$ operation is applied with the ancilla as control. A final Hadamard gate on the ancilla completes the interferometric readout, as shown in Fig.~\ref{fig:qae}. The overall transformation can be written as
\begin{equation}\label{eq:qae_stateprime}
\ket{\psi'}
=
(\mathrm{H}_a\otimes I)\,(\mathrm{C}\!-\!F_1)\,(\mathrm{H}_a\otimes I)\big[\ket{0_a}\otimes V\ket{\psi_0}\big].
\end{equation}
The expectation value $\langle F_1\rangle$ is then extracted from the ancilla statistics via Eq.~\eqref{eq:qae_prob}. 
To further reduce the circuit depth, we variationally compress the full interferometric circuit into {another} parameterized circuit $V'$. Denoting the input state by $\ket{\Phi_0}=\ket{0_a}\otimes\ket{\psi_0}$, 
the compressed circuit is trained so that
\begin{equation}
V'\ket{\Phi_0}
\approx
(\mathrm{H}_a\otimes I)\,(\mathrm{C}\!-\!F_1)\,(\mathrm{H}_a\otimes I)\big[\ket{0_a}\otimes V\ket{\psi_0}\big].
\end{equation}
The results shown in Figs.~\ref{fig:noise}, \ref{fig:figure2}, \ref{fig:dy}, and \ref{fig:dis} are obtained using this compressed readout workflow. For results reported in Fig.~\ref{fig:noise}, the circuit structure is $V^{\mathrm{Identity}}V'$.

\section{Finite-size ground-state diagram under disorder and imaginary fields}\label{suppphase}
In the main text, we consider the disordered Wen--plaquette model,
\begin{equation}\label{suppdis}
H_{\rm dis}
= H_{\rm wen}
- g\sum_{i}\sigma^{x}_{i}
- \sum_{i}h_{i}\sigma^{z}_{i},
\end{equation}
where the transverse-field deformation is
\begin{equation}\label{suppH}
H_{X}
= H_{\rm wen}
- g\sum_{i}\sigma^{x}_{i},
\end{equation}
and the underlying plaquette Hamiltonian is
\begin{equation}\label{suppwen}
H_{\rm wen}
= -J\sum_{i}F_{i},
\qquad
F_{i}=\sigma_{i}^{x}\sigma_{i+\hat{x}}^{y}\sigma_{i+\hat{x}+\hat{y}}^{x}\sigma_{i+\hat{y}}^{y}.
\end{equation}
Here, for notational simplicity, each plaquette is labeled by a single index $i$; the corresponding site and plaquette labeling convention is shown in Fig.~1(b) of the main text.

\begin{figure}[h]
    \centering
    \includegraphics[width=0.7\linewidth]{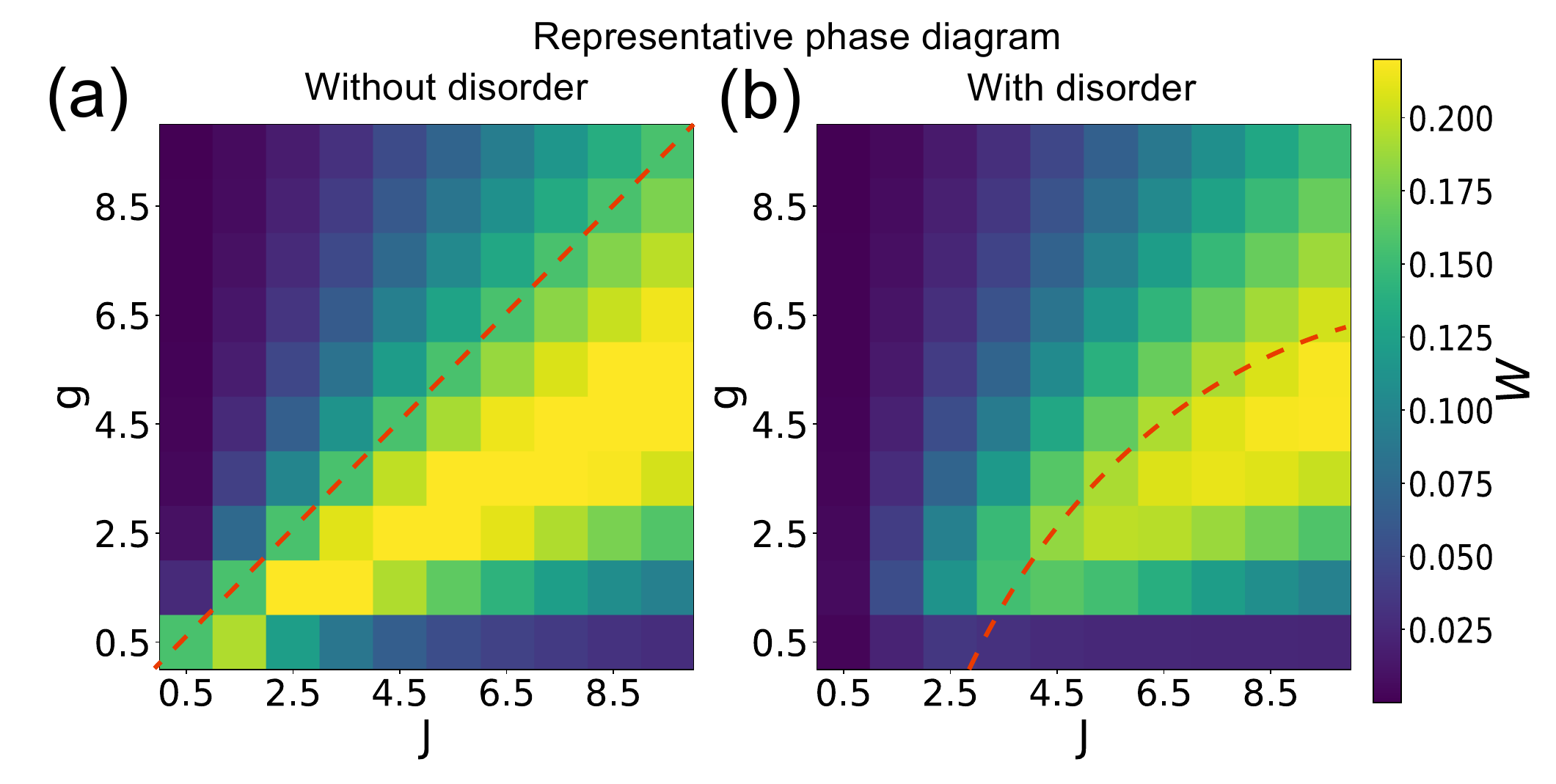}
    \caption{Classically estimated ground-state diagram of the disordered Wen--plaquette model [Eq.~\ref{suppdis}], obtained by exact diagonalization. The diagnostic $W$ is the large Wilson loop defined in Eq.~\ref{w}. Panels \textbf{(a)} and \textbf{(b)} correspond to disorder strengths $h=1$ and $h=5$, respectively. }
    \label{fig:phasediagram_x}
\end{figure}

The corresponding finite-size response structure obtained from exact diagonalization is shown in Fig.~\ref{fig:phasediagram_x}. Upon introducing disorder, the Wilson-loop diagnostic develops a broadened crossover boundary that separates a regime with appreciable nonlocal loop response from a disorder-dominated regime. At the same time, the overall organization of the finite-size response map remains qualitatively similar to that of the clean system. These results indicate that the finite-size stabilizer and loop diagnostics of the Wen--plaquette model remain robust against moderate on-site disorder.

\begin{figure}[h]
	\centering
	\includegraphics[width=0.7\linewidth]{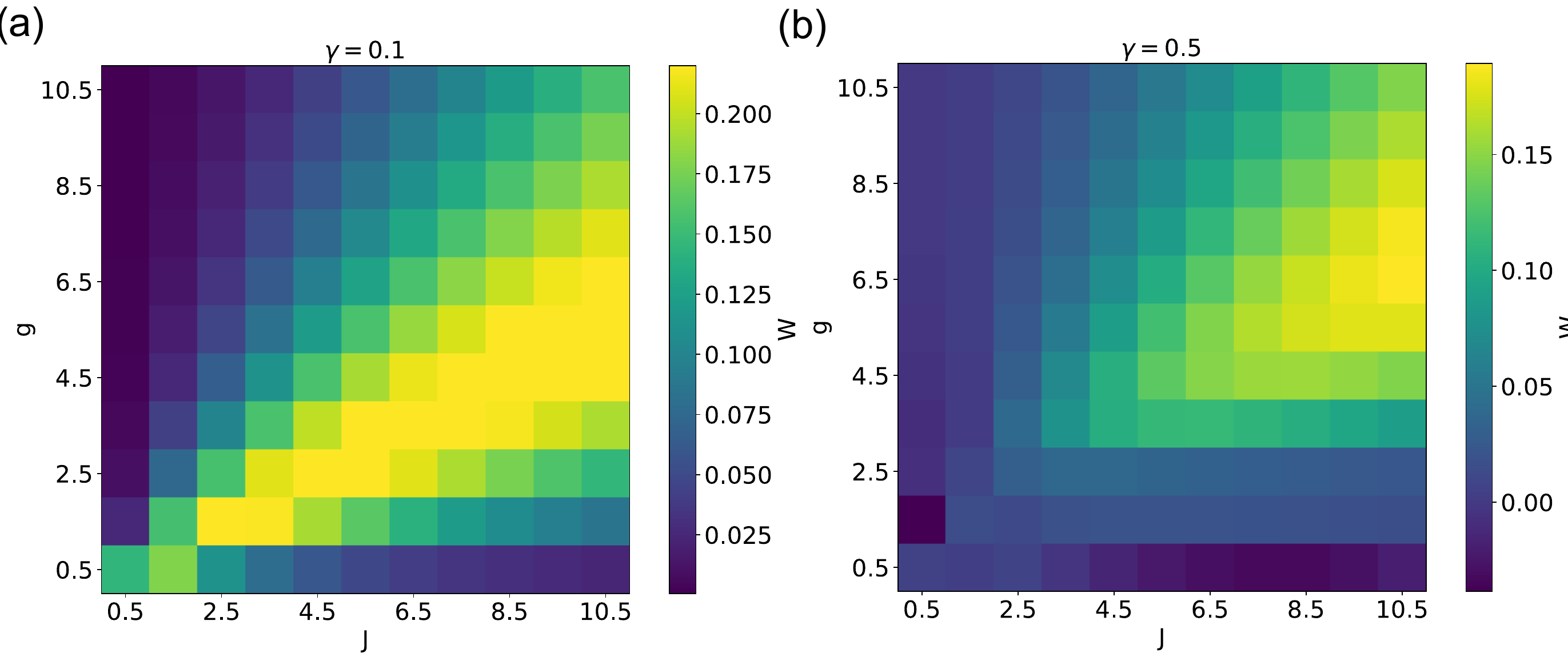}
	\caption{Classically estimated ground-state diagram of the non-Hermitian {Wen--plaquette} model [Eq.~\ref{apnon}], obtained by exact diagonalization. The diagnostic $W$ is the large Wilson loop defined in Eq.~\ref{w}.  Panels  {(a)} and  {(b)} correspond to the strength of imaginary fields: $\gamma=0.1$ and $\gamma=0.5$, respectively. Throughout, we fix $J=5$ and $g=1$.}
	\label{fig:phase}
\end{figure}

We next examine a non-Hermitian extension that incorporates an effective onsite nonunitary loss,
\begin{equation}\label{apnon}
H_{\gamma}=H_{\rm wen}-g\sum_{i}\sigma^{x}_{i}+i\gamma\sum_{i}\sigma^{z}_{i}.
\end{equation}
To characterize the associated finite-size response structure, we select the right eigenstate $\ket{\psi_{\rm R}}$ whose eigenvalue has the smallest real part and evaluate observables using the normalized right-state expectation value.
In particular, the Wilson-loop response shown in Fig.~\ref{fig:phase} is computed from $\langle W\rangle_{\rm R}$ under this convention. The resulting response map closely resembles that obtained for the disordered Hermitian model in Fig.~\ref{fig:phasediagram_x}: as $\gamma$ increases, the nonlocal loop signal is progressively suppressed, in close analogy with the effect of increasing disorder strength $h$. This correspondence suggests that, at the level of the finite-size Wilson-loop diagnostic studied here, onsite non-Hermitian fields and static disorder disrupt the plaquette structure in qualitatively similar ways.

\section{Quantum hardware}

\begin{figure}[h]
    \centering
    \includegraphics[width=0.8\linewidth]{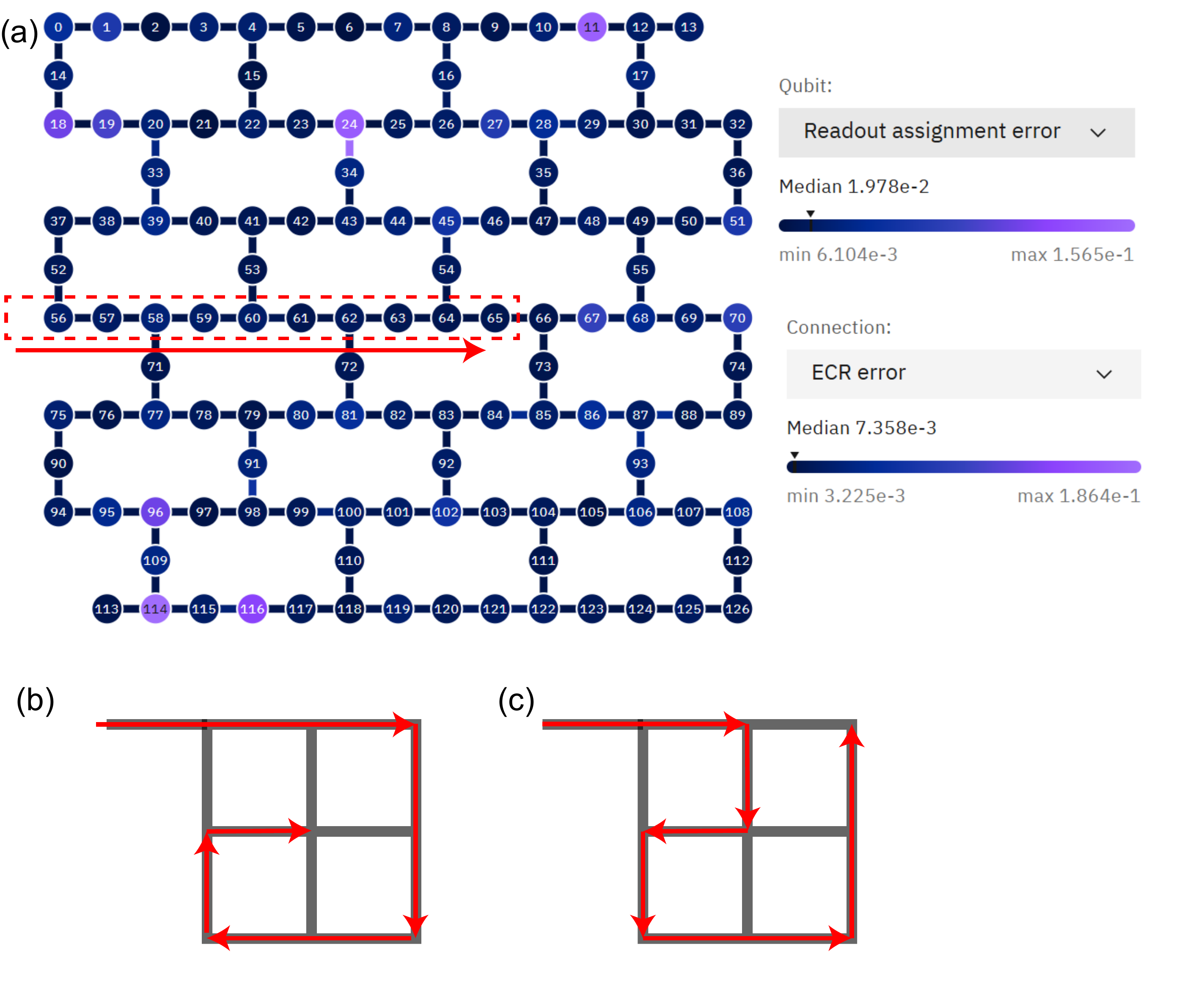}
    \caption{Layout of the IBM \texttt{ibm\_brisbane} processor (127 qubits). (a) In this work, we use the linear chain of physical qubits [56, \ldots, 66] for our simulations. For the ancilla-assisted interferometric readout protocol, qubit 56 serves as the ancilla, while the remaining qubits act as system qubits. \textbf{(b,c)} Mapping from the physical qubit chain to the effective square-lattice geometry of the target model. \textbf{(b)} Configuration used to measure the large Wilson loop. \textbf{(c)} Configuration used to measure the plaquette operator $F_{1}$ (see Fig.~9 in Appendix).}
    \label{fig:ibm}
\end{figure}

\begin{figure}
    \centering
    \includegraphics[width=0.5\linewidth]{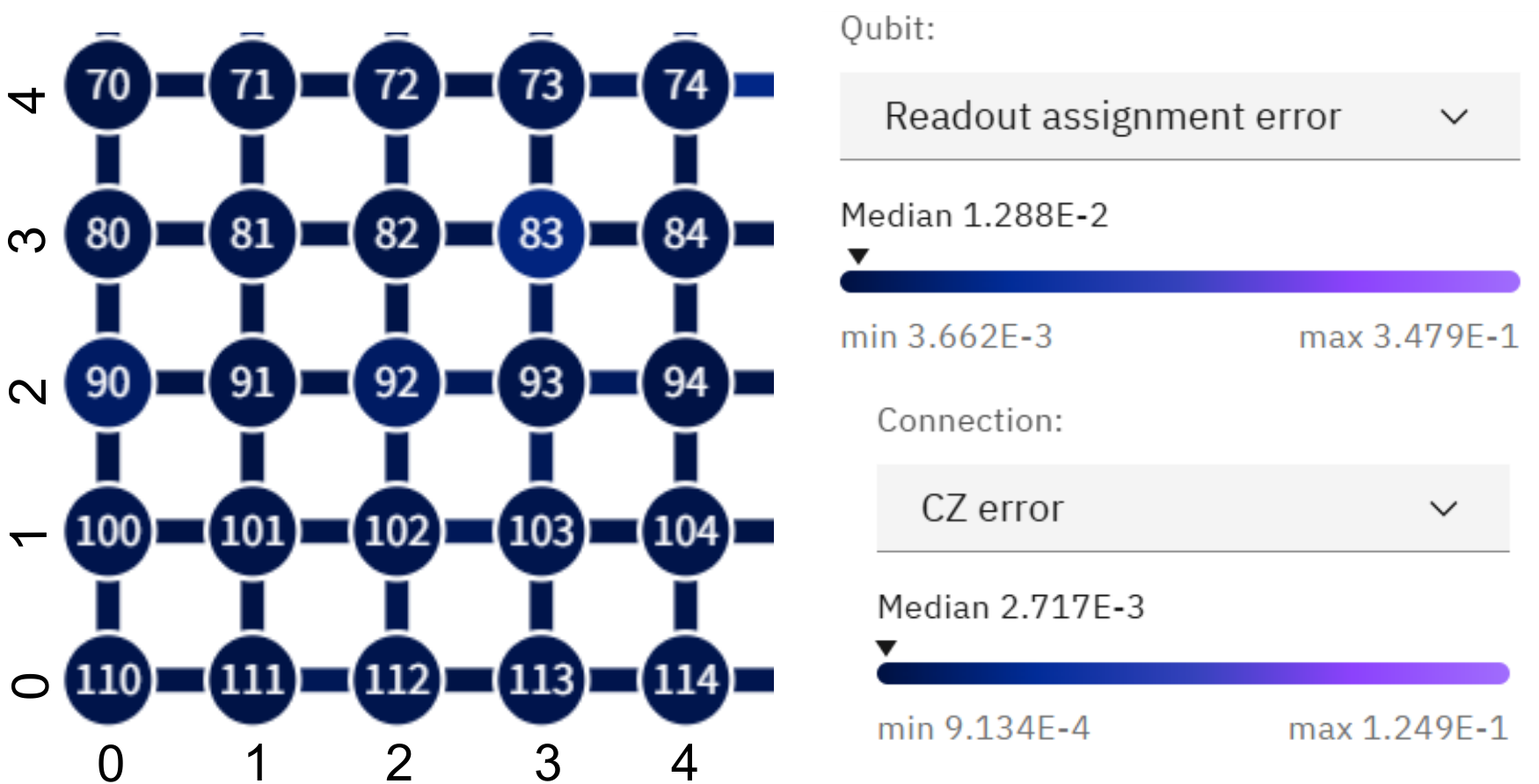}
    \caption{Layout and the error information of the 2D IBM Miami device.}
    \label{fig:2dibm}
\end{figure}

We present hardware data obtained on the IBM Quantum \texttt{ibm\_brisbane} processor, which comprises 127 qubits, as shown in Fig.~\ref{fig:ibm}. For the simulations reported here, we use the qubit chain $[56,\ldots,66]$, selected for its comparatively low noise levels. Along this chain, the average error rate of the echoed cross-resonance (ECR) gates is approximately $0.5\%$. This backend is used for the simulations shown in Figs.~3--6 of the main text. For the results presented in Fig.~7 of the main text, we instead use the two-dimensional IBM Quantum \texttt{ibm\_miami}  {NightHawk} device; additional details are provided in Fig.~\ref{fig:2dibm}.

\section{Large-scale ground-state preparation protocols}
\subsection{Ancilla-assisted projector implementation (block encoding, LCU, and feedforward)}
In this section, we propose a scalable protocol for preparing the Wen--plaquette ground-state sector. A representative state in the clean Wen--plaquette ground-state sector has the simple stabilizer form
\begin{equation}\label{g}
\ket{G}\propto \prod_{i}(I+F_{i})\ket{0}^{\otimes N},
\end{equation}
where $N$ is the number of qubits and $I$ denotes the identity operator. Here, for notational simplicity, each plaquette operator is labeled by a single index $i$.

This construction explicitly enforces the $+1$ stabilizer constraint on every plaquette. Indeed, for any $j$,
\begin{equation}
F_{j}\prod_{i}(I+F_{i})\ket{0}^{\otimes N}
=
F_{j}(I+F_{j})\prod_{i\neq j}(I+F_{i})\ket{0}^{\otimes N}
=
(I+F_{j})\prod_{i\neq j}(I+F_{i})\ket{0}^{\otimes N}
=
\prod_{i}(I+F_{i})\ket{0}^{\otimes N},
\end{equation}
where we used $F_j^2=I$ and $[F_j,F_i]=0$ for $i\neq j$. Hence $\ket{G}$ satisfies
\begin{equation}
F_j\ket{G}=\ket{G}
\end{equation}
for all plaquettes and is therefore a simultaneous $+1$ eigenstate of the full stabilizer set.

To prepare this state on quantum hardware, one must implement the nonunitary projector $(I+F_i)/2$ on each plaquette. Below, we outline several approaches.

First, the nonunitary projector can be implemented via a unitary dilation (block encoding). Concretely, one constructs a unitary $U_i$ acting on the four plaquette qubits and an ancilla such that the desired map appears as a block in the ancilla-$\ket{0}$ subspace. Preparing the ancilla in $\ket{0}$, applying $U_i$, and post-selecting the ancilla outcome $\ket{0}$ implements the projector $(I+F_i)/2$ on the system.

This unitary-dilation viewpoint is closely connected to the linear-combination-of-unitaries (LCU) framework. The stabilizer projector can be written as
\begin{equation}
\frac{I+F_i}{2}
=
\frac{1}{2}\big(U_0+U_1\big),
\qquad
U_0=I,\quad U_1=F_i.
\end{equation}
In the LCU implementation, an ancilla qubit is prepared in $\ket{+}=(\ket{0}+\ket{1})/\sqrt{2}$ and used to coherently select between the two branches: conditioned on the ancilla state, one applies either $U_0$ or $U_1$, equivalently a controlled-$F_i$. Measuring the ancilla in the $\{\ket{+},\ket{-}\}$ basis and post-selecting the $\ket{+}$ outcome then realizes the nonunitary map $(I+F_i)/2$ on the system.

A more experimentally accessible strategy exploits mid-circuit measurements and real-time classical control. In this approach, the four qubits of plaquette $i$ are coupled to an ancilla initially prepared in $\ket{0}$. A short sequence of controlled-Pauli gates maps the eigenvalue of $F_i$ onto the ancilla:
\begin{equation}
\ket{\psi}\ket{0}
\longrightarrow
\frac{1}{2}\Big[(I+F_i)\ket{\psi}\Big]\ket{0}
+
\frac{1}{2}\Big[(I-F_i)\ket{\psi}\Big]\ket{1}.
\end{equation}

Measuring the ancilla gives an outcome $m\in\{0,1\}$.

If $m=0$, the system has been projected into the $+1$ stabilizer sector $\ket{\psi_{+}}$, thereby implementing $(I+F_i)/2$.

If $m=1$, the system is projected into the $-1$ stabilizer sector $\ket{\psi_{-}}$. One must then apply a corrective operator $C_i$ that anticommutes with $F_i$. For the Wen--plaquette stabilizer, this can be achieved by applying a single-qubit Pauli $\sigma^z$ on any one of the four plaquette qubits, which flips the eigenvalue of $F_i$.

Technically, this method requires real-time classical feedforward, meaning that the corrective operation must be conditioned explicitly on the measurement outcome. Compared with the block-encoding or LCU-based constructions discussed above, this introduces additional experimental overhead and may therefore be more challenging to implement on hardware with limited mid-circuit control capabilities.

While conceptually straightforward, implementing this approach on present-day IBM Quantum processors faces architectural limitations. In superconducting-qubit devices, restricted qubit connectivity makes it difficult to perform the required multi-qubit stabilizer measurements without extensive SWAP operations, which rapidly increase circuit depth and noise.

By contrast, trapped-ion platforms naturally provide all-to-all connectivity and long coherence times, making them well suited to this preparation scheme. Their ability to execute arbitrary multi-qubit interactions and mid-circuit feedback makes them particularly promising for realizing large-scale topological states following the protocol outlined above.

\subsection{Coherent representative-qubit method}\label{rep}

To avoid mid-circuit measurements and postselection in the hardware demonstration, we employ the coherent representative-qubit construction introduced in Ref.~\cite{xu2023digital}. The method generates a state in a prescribed plaquette-stabilizer sector through unitary controlled operations. It realizes the state-construction identity
\begin{equation}
\ket{\Phi_i^{\pm}}
\propto
(\mathbb{I}\pm F_i)\ket{0}_r\ket{\psi},
\end{equation}
provided that the representative qubit \(r\) is initialized in \(\ket{0}_r\). It should therefore be distinguished from a deterministic implementation of the nonunitary projector
\(\Pi_i=(\mathbb{I}+F_i)/2\) acting on an arbitrary many-body input state.

For a plaquette \(i\), we choose one of its four corner qubits as the representative qubit \(r\) and decompose the stabilizer as
\begin{equation}
F_i=P_r\,\widetilde F_i,
\qquad
P_r\in\{\sigma_r^x,\sigma_r^y\},
\end{equation}
where \(P_r\) acts on the representative qubit and \(\widetilde F_i\) is the product of the remaining three Pauli operators. For example, for
\begin{equation}
F_{i,j}
=
\sigma_{i,j}^{x}
\sigma_{i+1,j}^{y}
\sigma_{i+1,j+1}^{x}
\sigma_{i,j+1}^{y},
\end{equation}
choosing the upper-right corner \(r=(i+1,j+1)\) gives
\begin{equation}
P_r=\sigma_{i+1,j+1}^{x},
\qquad
\widetilde F_{i,j}
=
\sigma_{i,j}^{x}
\sigma_{i+1,j}^{y}
\sigma_{i,j+1}^{y}.
\end{equation}

We then define the representative-controlled Pauli string
\begin{equation}\label{eq:controlled_string}
C_r(\widetilde F_i)
=
\ket{0}\!\bra{0}_r\otimes\mathbb{I}
+
\ket{1}\!\bra{1}_r\otimes\widetilde F_i.
\end{equation}
Because the three Pauli factors in \(\widetilde F_i\) act on different qubits, this operation can be decomposed as
\begin{equation}
C_r(\widetilde F_i)
=
\prod_{q\in\partial i\setminus r}C_r(P_q^{(i)}),
\end{equation}
where \(P_q^{(i)}\in\{\sigma_q^x,\sigma_q^y\}\) is the Pauli factor acting on corner \(q\). At the circuit level, a controlled-\(\sigma^x\) operation is a CNOT gate, while a controlled-\(\sigma^y\) operation can be implemented as
\begin{equation}
C_r(\sigma_q^y)
=
S_q\,\mathrm{CNOT}_{r,q}\,S_q^\dagger,
\end{equation}
before transpilation into the native gate set of the selected processor.

If \(P_r=\sigma_r^x\), the representative qubit is prepared in
\begin{equation}
\ket{+_x}_r
=
H_r\ket{0}_r
=
\frac{\ket{0}_r+\ket{1}_r}{\sqrt{2}}.
\end{equation}
Applying Eq.~\eqref{eq:controlled_string} gives
\begin{align}
\ket{+_x}_r\ket{\psi}
&\xrightarrow{\,C_r(\widetilde F_i)\,}
\frac{1}{\sqrt{2}}
\left(
\ket{0}_r\ket{\psi}
+
\ket{1}_r\widetilde F_i\ket{\psi}
\right)
\nonumber\\
&=
\frac{1}{\sqrt{2}}
(\mathbb{I}+F_i)\ket{0}_r\ket{\psi},
\label{eq:directprep:repqubit_X}
\end{align}
where \(F_i=\sigma_r^x\widetilde F_i\) and
\(\sigma_r^x\ket{0}_r=\ket{1}_r\). The two branches are orthogonal because they contain different representative-qubit states, so the factor \(1/\sqrt{2}\) provides the correct normalization. The resulting state satisfies
\begin{equation}
F_i\ket{\Phi_i^+}=\ket{\Phi_i^+},
\end{equation}
and therefore lies exactly in the \(F_i=+1\) stabilizer sector.

If \(P_r=\sigma_r^y\), the representative qubit is instead prepared in
\begin{equation}
\ket{+_y}_r
=
S_rH_r\ket{0}_r
=
\frac{\ket{0}_r+i\ket{1}_r}{\sqrt{2}}.
\end{equation}
Using \(\sigma_r^y\ket{0}_r=i\ket{1}_r\), one obtains
\begin{align}
\ket{+_y}_r\ket{\psi}
&\xrightarrow{\,C_r(\widetilde F_i)\,}
\frac{1}{\sqrt{2}}
\left(
\ket{0}_r\ket{\psi}
+
i\ket{1}_r\widetilde F_i\ket{\psi}
\right)
\nonumber\\
&=
\frac{1}{\sqrt{2}}
(\mathbb{I}+F_i)\ket{0}_r\ket{\psi},
\label{eq:directprep:repqubit_Y}
\end{align}
which likewise satisfies \(F_i\ket{\Phi_i^+}=\ket{\Phi_i^+}\). The corresponding \(F_i=-1\) state can be generated by preparing \(\ket{-_x}\) or \(\ket{-_y}\), producing a state proportional to \((\mathbb{I}-F_i)\ket{0}_r\ket{\psi}\).

No measurement or probabilistic postselection is required because the representative qubit remains part of the final many-body state. However, the construction is not equivalent to applying \((\mathbb{I}+F_i)/2\) to an arbitrary state: its validity as a unitary state-preparation identity relies on the prescribed initialization of the representative qubit. Moreover, because neighboring plaquettes share qubits, a multi-plaquette implementation must specify the representative-qubit assignment and gate ordering. In the hardware protocol, we use the fixed lattice sweep shown in Fig.~7(a) and assess the final prepared sector by measuring all accessible plaquette expectations and their lattice average \(\bar F\), rather than interpreting each circuit block as an independent nonunitary projection.

In this work, the representative-qubit construction is used to prepare and verify the Wen--plaquette stabilizer-sector state employed in Fig.~7 of the main text. Its decomposition into local controlled-Pauli operations avoids variational optimization and provides a hardware-compatible route toward larger square-lattice implementations, with the required routing determined by the connectivity of the selected processor.

\subsection{Noise effect}\label{noise}
The scalability of the coherent representative-qubit construction is governed primarily by the compiled circuit depth. As illustrated in Fig.~\ref{fig:large2}(a), the plaquette constraints are imposed through a layered sweep across the lattice. Projector blocks within the same sweep stage may be executed in parallel provided that they do not share physical qubits. Thus, although the total number of plaquettes scales as $N_{\rm p}=(L_x-1)(L_y-1)$, the number of ideal sweep stages grows only with the linear dimensions of the lattice, approximately as $N_{\rm sweep}\sim L_x+L_y-3$, up to hardware-dependent scheduling and routing factors. The actual entangling-gate depth must therefore be determined from the transpiled circuit for the chosen processor.

For each plaquette, the representative qubit is directly connected to two neighboring corners, whereas the diagonally opposite corner is reached through an intermediate qubit. After the required local Clifford basis rotations, the resulting three-target controlled operation is compiled into four physical $\mathrm{CZ}$ gates arranged in three local entangling layers. For the $5\times5$ implementation, the circuit contains 16 plaquettes, and the compiled preparation used here has
\begin{equation}
N_{\rm p}=16,
\qquad
N_{\rm CZ}=64,
\qquad
D_{\rm CZ}=48.
\end{equation}
Here, $N_{\rm CZ}$ is the total number of physical $\mathrm{CZ}$ gates, whereas $D_{\rm CZ}$ is the number of sequential $\mathrm{CZ}$ layers after parallel scheduling. The former determines the accumulated exposure to entangling-gate errors, while the latter controls the entangling-gate contribution to the circuit duration and decoherence.

To obtain an order-of-magnitude estimate of the resulting attenuation, we treat gate errors and decoherence independently. Under a two-qubit depolarizing approximation, an average $\mathrm{CZ}$ infidelity $\bar{\epsilon}_{\rm CZ}$ produces the Pauli attenuation factor
$\lambda_{\rm CZ}\simeq1-4\bar{\epsilon}_{\rm CZ}/3$. The lattice-averaged plaquette response may then be estimated as
\begin{equation}\label{eq:compiled_noise_estimate}
\frac{\overline{F}_{\rm hw}}
{\overline{F}_{\rm ideal}}
\simeq
\lambda_{\rm CZ}^{\,N_{\rm CZ}^{\rm cone}}
\exp\left[
-\frac{
D_{\rm CZ}\tau_{\rm CZ}
+
D_{\rm 1q}\tau_{\rm 1q}
}{T_{\rm eff}}
\right],
\end{equation}
where $N_{\rm CZ}^{\rm cone}$ denotes the number of compiled $\mathrm{CZ}$ gates within the effective causal cone of the measured response. As a conservative estimate, we take $N_{\rm CZ}^{\rm cone}=N_{\rm CZ}=64$.

Using the representative calibration values
$\bar{\epsilon}_{\rm CZ}=2.5\times10^{-3}$,
$\tau_{\rm CZ}=68~\mathrm{ns}$, and
$T_{\rm eff}=100~\mu\mathrm{s}$, the entangling-gate attenuation is
$\lambda_{\rm CZ}^{64}\simeq0.808$. The sequential $\mathrm{CZ}$ layers contribute a duration
$D_{\rm CZ}\tau_{\rm CZ}=3.264~\mu\mathrm{s}$, corresponding to a coherence factor of approximately $0.968$. Neglecting the smaller single-qubit contribution therefore gives
\begin{equation}
\frac{\overline{F}_{\rm hw}}
{\overline{F}_{\rm ideal}}
\simeq
0.808\times0.968
\simeq
0.78,
\end{equation}
which fits our measured data shown in Fig.~\ref{fig:large2} (main text).
This simplified estimate is consistent with the experimentally observed visibility of the plaquette response on the $5\times5$ lattice. Nevertheless, it shows quantitatively that extending the construction to larger lattices will require lower two-qubit error rates, reduced routing overhead, and more efficient parallel scheduling.
\newpage

\flushbottom
\end{document}